\documentclass[12pt]{article} 
\pdfoutput=1
\usepackage[hyperfootnotes=false]{hyperref}
\usepackage{epsfig}
\usepackage{amsmath}
\usepackage{amssymb}
\usepackage{graphicx}
\usepackage{float}
  \newlength{\abstractwidth}
\usepackage{comment}

\usepackage{mciteplus} 
\usepackage{subcaption}

  \newcommand{\be}{\begin{equation}}
  \newcommand{\bea}{\begin{eqnarray}}
  \newcommand{\eea}{\end{eqnarray}}
  \newcommand{\beq}{\begin{equation}}
  \newcommand{\ee}{\end{equation}}
  \newcommand{\eeq}{\end{equation}}

	\newcommand{\eqn}[1]{\begin{equation}\begin{split} #1 \end{split}\end{equation}}
	
	\newcommand{\lp}{\left (}
	\newcommand{\rp}{\right )}

	\newcommand{\pa}[2]{\frac{\partial #1}{\partial #2}}
	\newcommand{\pd}{\partial}

	\newcommand{\tr}{\text{tr} \,}

	\newcommand{\hf}{\frac{1}{2}}

	\newcommand{\lb}{\left [}
	\newcommand{\rb}{\right ]}

	\newcommand{\ev}[1]{\left \langle #1 \right \rangle}

\newcommand{\lt}[1]{\ev{ \tr #1} }

\def\la{\label}

\def\nref#1{(\ref{#1})}

\begin{document}

\begin{titlepage}
  \bigskip

  \bigskip\bigskip

  \bigskip

\begin{center}
 
\centerline
{\Large  {Bootstraps to Strings:}}
\bigskip
{\Large  Solving Random Matrix Models with Positivity} 
 \bigskip

 \bigskip
{\Large \bf { }} 
    \bigskip
\bigskip
\end{center}

  \begin{center}

	   {Henry W. Lin}
  \bigskip \rm
  
\bigskip
 Jadwin Hall, Princeton University, Princeton, NJ 08540, USA\\
 Google, Mountain View, CA 94043, USA\\

  \bigskip \rm
\bigskip
 
\rm

\bigskip
\bigskip

  \end{center}

 \bigskip\bigskip
  \begin{abstract}

	  A new approach to solving random matrix models directly in the large $N$ limit is developed. First, a set of numerical values for some low-pt correlation functions is guessed. The large $N$ loop equations are then used to generate values of higher-pt correlation functions based on this guess. Then one tests whether these higher-pt functions are consistent with positivity requirements, e.g., $\langle \text{tr }M^{2k} \rangle \ge 0$. If not, the guessed values are systematically ruled out.
In this way, one can constrain the correlation functions of random matrices to a tiny subregion which contains (and perhaps converges to) the true solution. This approach is tested on single and multi-matrix models and handily reproduces known solutions. It also produces strong results for multi-matrix models which are not believed to be solvable.
A tantalizing possibility is that this method could be used to search for new critical points, or string worldsheet theories.

 \medskip
  \noindent
  \end{abstract}
\bigskip \bigskip \bigskip

\vspace{1cm}

\begin{figure}[H]
\hspace*{-1in}
\begin{raggedleft}
	\includegraphics[scale=0.25, trim = -770 0 0 0]{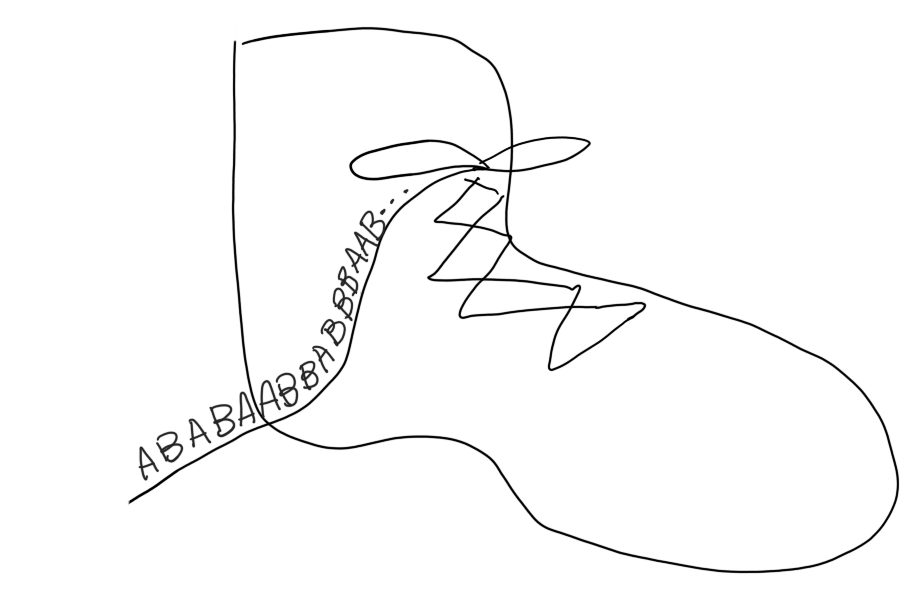}
\end{raggedleft}
\end{figure}

\vspace{2cm}
\end{titlepage}

   \tableofcontents
   \section{Introduction} 
\def\O{\mathcal{O}}

From ancient days, sages appreciated that certain large $N$ theories simplify dramatically.  
For matrix theories in the 't Hooft limit, one needs to sum only a tiny subset of all possible Feynman diagrams, the planar ones \cite{tHooft:1973jz, 'tHooft:1974hx}. 
  However, with a few notable exceptions, this simplification is not enough: even in the 't Hooft limit, most matrix theories are impossible to solve. This is true even for zero dimensional statistical ensembles of a small number of matrices. With the exception of the single matrix model \cite{wigner1993characteristic, bipz, Bessis:1980ss,douglas1990strings}, almost  
  all models remain unsolved\footnote{See \cite{Kazakov:2000aq, eynard2011formal} for a list of solvable models and a more comprehensive review of the techniques.}.

  In this paper, we propose a method to solve multi-matrix models in the strict large $N$ limit. For our purposes, a 2-matrix model is defined by an integral of the form 
  \eqn{Z = \lim_{N \to \infty} \int dA \, dB\, e^{- N \, \text{Tr } V(A,B)},}
where the integration measure is the uniform measure over Hermitian matrices, and $V$ is a polynomial, the coefficients of which will be considered as couplings. Here, ``solving'' a matrix model means (a) determining what values of the couplings for which the integral exists and (b) computing $Z$ as well as all possible single trace expectation values when they exist as a function of the couplings, to leading order in $1/N$:
\eqn{\ev{\text{Tr } \mathcal{O}(A,B)} = \lim_{N \to \infty} {1 \over Z} \int dA \, dB e^{- N\, \text{Tr } V(A,B)} \text{Tr } \mathcal{O}(A,B),}
where $\O$ is some arbitrary polynomial in the matrices. Unlike in the finite $N$ case, even step (a) is highly non-trivial in most cases.

A complementary tool for studying such systems is numerics, e.g., Monte Carlo simulation. A downside of this approach is that the large $N$ limit is numerically difficult; one needs to simulate ever more degrees of freedom, even though the underlying physics is simplifying. 
A more conceptual problem is that some matrix ensembles are ill-defined at finite $N$ but are well-defined at infinite $N$. 
Far from a mere technicality, such situations are usually\footnote{Another case is when multi-cuts merge \cite{Klebanov:2003wg}.} what is required to compare with continuum quantum gravity calculations (see \cite{DiFrancesco:1993cyw} for a review).
In this paper, we will instead develop a method that works well at infinite $N$ and strong (or weak) coupling. The method involves both analytics and numerics and deploys a philosophy similar to the conformal bootstrap (see \cite{Anderson:2016rcw, Jevicki:1983hb} for an approach to lattice field theory that is morally quite similar). 
The analytical part involves writing down the loop equations and finding simple positivity relations on matrix correlators. Then one numerically searches over the space of possible values of the correlation functions which are consistent with the analytics. 
The resulting bounds obtained are rigorous, even if they were obtained with numerics.

A word about notation: we normalize little trace and big Trace by $\tr \mathbf{1} = {1 \over N} \text{Tr } \mathbf{1}=  1$ and will often denote single matrix correlators $t_k = \ev{\tr A^k}$. This paper is organized as follows. In Section \ref{loopS}, we review the equations of motion for (multi-)matrix models, known as the ``loop equations.'' In Section \ref{posS}, we discuss positivity constraints on correlation functions $\ev{\tr \O(A,B)}$. As we will emphasize, an arbitrary list of numbers will in general not be a consistent set of correlation functions. In Section \ref{bootS}, we illustrate how the method works by reproducing known exact solutions and move on in Section \ref{bootMS} to solve a model which is not known to be integrable. We discuss some open questions in Section \ref{disS}.



\section{The Loop Equations \la{loopS}}

The loop equations of a random matrix theory are nothing but the Schwinger-Dyson equations. They are derived by integrating a total derivative. For a single matrix model,
\eqn{0 = \int dM {\pd \over \pd M_{ij}} \lp (M^k)_{ij} e^{-S(M)}\rp \la{loopeqn}.}
The derivative can act on either the $M^k$ or the $e^{-S}$ term, giving
\eqn{\ev{\tr M^k V'(M)} = \sum_{\ell=0}^{k-1} \ev{\tr M^{\ell}} \ev{\tr M^{k-\ell -1}}.}
In writing the double trace term as a product of single traces, we have used large $N$ factorization.
These equations also have a diagrammatic interpretation. Set $V(M) = \hf M^2 + {g \over 3} M^3$ for simplicity. Then consider the computation of a $k$-pt function. At large $N$, this is a sum of planar diagrams with $k$ external lines (we will not need 't Hooft's double line notation). Choose one of the external lines and follow it into the blob:

\vspace*{-0.2cm}
\begin{figure}[H]
\begin{center}
\includegraphics[scale=0.3]{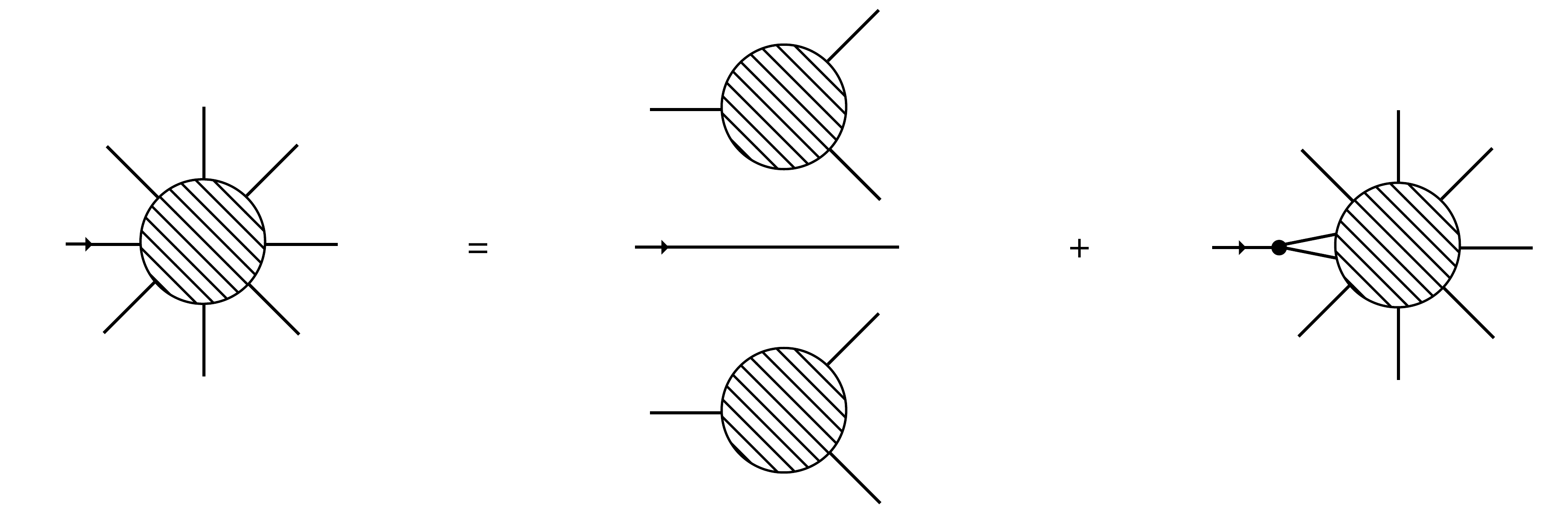}
\label{loop-fig}
\end{center}
\end{figure}
\vspace*{-0.6cm}
\noindent 
There are two possibilities. If this edge never encounters a vertex, it must become another external line. In this case, it divides the planar diagrams into two parts. Otherwise, the line must end in a vertex. So we get a relationship between lower-pt correlation functions and higher-pt ones.

The diagrammatic interpretation of the loop equations is particularly useful when we consider more complicated multi-matrix models. 
One can easily read off the form of the loop equations without any calculation. If we consider the trace of a monomial of degree $D$ in the matrices, we can follow each external line into the blob, and hence derive $D$ different loop equations. (Some of these may be redundant due to the cyclicity of the trace or other symmetries.)
Alternatively, we could consider 
\eqn{0 = \int_{A,B} {\pd \over \pd A_{ij}} \lb \O(A,B)_{ij} e^{-S(A,B)}\rb = \int_{A,B}  {\pd \over \pd B_{ij}} \lb \O'(A,B)_{ij} e^{-S(A,B)}\rb 
\la{loopeqnmulti} 
}
Here $\O, \O'$ are arbitrary monomials of $A$ and $B$.
Notice that in this expression, $\O$ is not a trace, so for example, we get different equations if we consider $\O = A^2B^2$ and $\O = AB^2A$.

\subsection{The search space}
As we have seen, the loop equations relate higher-pt functions to lower-pt functions. A natural question is: what is the minimum number of correlators $s_*$ that we need to know in order to determine the rest? 
For reasons that will soon be apparent, we call a minimal subspace of correlators $S$ the ``search space.'' If the values of the correlators in the search space are known, the rest of the correlators follow. By definition, $\dim{S} = s_*$. While the value of $s_*$ is unique, there may be many possible choices for $S$. 
One should not confuse $s_*$ with the number of valid solutions to the loop equations. In general, the number of valid solutions is less than $s_*$ because of positivity constraints; see Appendix \ref{exactA} for a precise analysis of the 1-matrix case. 
The upshot is that for a potential of degree $d$, $s_* = d-2$.  This is the same as the number of formal solutions to the loop equations, e.g., solutions where the eigenvalue density does not have to be positive. 

An immediate question for the multi-matrix model is how to estimate $s_*$. Since the number of correlation functions of fixed degree is growing exponentially for even a 2-matrix model, it may not seem obvious that $s_*$ should even be finite.
In fact, we will argue that if we know all correlation functions of degree at most $k_*$, then the loop equations determine the rest. Suppose we consider a general matrix model with $m$ matrices, and a potential which is a polynomial of degree $D$ in the $m$ matrices. 
 A crude estimate of $k_*$ can be obtained as follows. There are roughly $m^k/k$ correlation functions of degree $k$. Each correlator gives approximately $k$ loop equations. So the number of loop equations for correlators of degree $k$ is $\sim m^k$. These loop equations will produce correlators of degree $k+D$. So we expect that when $k$ gets large enough so that $m^k \sim m^{k+D} / (k+D)$, we will have enough equations to determine the rest of the correlators. This occurs at $k_*  \sim m^D $. 

In practice, this crude estimate is pessimistic. For the simple case $m=2$, $D=3$, direct calculation gives $s_* \le  5$, e.g. knowing the 5 traces $A,B, A^2, AB, B^2$ is enough to determine the rest of the correlation functions. If the model has $A\leftrightarrow B$ symmetry, just knowing the 3 correlation functions $A, A^2, AB$ is enough. Additional symmetries will typically simplify both the loop equations and reduce the value of $s_*$.

To check this, it is rather straightforward to write a simple program that will generate all the loop equations up to a fixed degree. The main subtlety is to ensure that only inequivalent traces are introduced at each new degree.
First list all correlation functions at a given degree. These consist of all words in the two variables $A$ and $B$, modulo the constraints:
\eqn{\tr x_1 x_2 \cdots x_k &= \tr x_2 \cdots x_k x_1,\\
	(\tr x_1 x_2 \cdots x_k)^* &= \tr x_k x_{k-1} \cdots x_1,\\
	\tr g(x_1 \cdots x_k) &= \tr x_1 \cdots x_k.
}
Here $g \in G$ is some global symmetry of the model, which in this case is just the $\mathbb{Z}_2$ symmetry $g(A) = B$, $g(B) = A$.
 More formally, if we have $m$ matrices, $\mathcal{A} = \{A_1, A_2 \cdots A_m\}$ then we are interested in the equivalence classes
\eqn{\O_{k,m} =  \mathcal{A}^k/(D_k \times G),}
where $D_k$ is the dihedral group. Here the dimension of $\O_{k,m}$ is the number of observables.

In models with low-degree polynomial potentials, there is often the additional symmetry $x_i \to x_i^T$. In this case all correlation functions are real. If $G = \{1\}$, the inequivalent traces are then in one-to-one correspondence with so-called {\it bracelets} in the combinatorics literature. The number of bracelets of length $k$, for $k$ even is
\eqn{B_m(k) = {1 \over k} \sum_{d|k} \phi(d) m^{k/d} + {1 \over 4} (m+1)m^{k/2} }
	where $\phi$ is the Euler totient function, and the sum is over all divisors of $k$. The main point of this formula is that a precise counting is somewhat complicated, which makes a precise estimate of $k_*$ difficult, but the large $k$ behavior is just $B \sim {1 \over k}  m^k$ as $k$ gets large.
An obvious question for future work is to understand if there is a simple criteria for calculating $s_*$ for a polynomial interaction in the matrices.\footnote{It is possible that the crude counting argument fails if there are significant degeneracies in the loop equations, so that the search space is infinite dimensional. However, so long as the number of new unknowns grows sufficiently slowly as the degree of the correlator increases, the constraints from positivity could be strong enough to overcome this growth. Since the number of positivity constraints grows exponentially with degree, a mild increase in the number of new unknowns with degree is unlikely to be a fundamental obstruction to the method. }
	\section{Positivity Constraints \la{posS} }
   Suppose one consults an oracle and receives a list of numbers which are purportedly the single-trace correlation functions of a matrix model. Here we ask the question: what consistency conditions does this list have to satisfy? In this section, we address this question. For the rest of this paper, we will restrict ourselves to the strict large $N$ limit. An incomplete discussion of positivity for $1/N$ corrections is relegated to Appendix \ref{1overN}.

\subsection{Positivity for one matrix ensembles}
We will consider the positivity constraints that can be derived from
\eqn{ \ev {\tr  \phi^\dagger \phi} \ge 0 \la{masterpos}}
Here $\phi$ is an arbitrary superposition of matrices; for the 1-matrix model, $\phi = \sum_k \alpha_k M^k$. This condition is equivalent to the following statement: if we consider the matrix $\mathcal{M}_{ij} = \ev{\tr M^{i+j}}$, all of its eigenvalues must be non-negative $\mathcal{M} \succeq 0$. In practice, we cannot enforce all of the constraints that follow from this condition, and we must choose a set of weaker constraints. Denoting the single trace correlators by $t_k = \ev{\tr M^k},$ we may impose 
\eqn{t_{2k} \ge 0, \qquad k \in \mathbb{Z}^+. \la{trivpos} }
These weaker constraints are linear in the single-trace correlators. We may also derive non-linear constraints from the above equation. For example, we can enforce positivity of a sub-matrix of $\mathcal{M}$:
\eqn{\mathcal{M}_{jk} = \begin{bmatrix}t_{2j} & t_{j+k} \\ t_{j+k} & t_{2k}\end{bmatrix} \succeq 0.\la{2matpos}}
The eigenvalues of this submatrix are $2\lambda_{j,k} = t_{2 j}+t_{2 k} \pm\sqrt{\left(t_{2 j}-t_{2 k}\right){}^2+4 t_{j+k}^2}$, which gives
\eqn{t_{2j}t_{2k} \ge t_{j+k}^2. \la{2pos} }
Notice that this inequality follows from just $\det \mathcal{M}_{jk} \ge 0$.
For $j=0$ this inequality has a simple interpretation. If we consider drawing eigenvalues randomly from the eigenvalue distribution $\rho(\lambda)$ of $M$, the inequality just says that the variance of the random variable $\lambda^k$ is non-negative. Furthermore, the $j=0$ constraint implies \nref{trivpos}. In writing \nref{2matpos}, we take all $t_k$ to be real, since $M$ is Hermitian.

The constraints \nref{trivpos} and \nref{2pos} came from considering $1 \times 1$ and $2 \times 2$ submatrices, respectively. In general, if we enforce positivity of $d \times d$ submatrices of $\mathcal{M}$, we will get a polynomial of degree $d$ constraint on the $t_k$ variables. These constraints will include statements such as ``all even moments of the random variable $X_k = (\lambda^k - t_k)$ are non-negative.'' These are essentially the inequalities that result from positivity of the eigenvalue distribution $\rho(\lambda)$ of $M$.

Note that we can find the boundary of allowed regions by finding the roots of the determinant of various sub-matrices $\det \mathcal{M}_{d\times d} = 0$. We can check that a matrix is non-negative by checking that the determinant of all upper-left submatrices are non-negative.

\def\M{\mathcal{M}}

\subsubsection{Relation to the Hamburger moment problem}
Large $N$ positivity in the single matrix model is closely related to positivity requirements on the moments of a real random variable. This is the subject of the {\it Hamburger moment problem}, which we now review. Given a sequence of real numbers $T = \{t_0=1, t_1, t_2 \cdots \}$, Hamburger asked for necessary and sufficient conditions on $T$ such that $t_k = \int_{-\infty}^{\infty} \rho(x) x^k $ for some positive measure $\rho$. The solution is that such a distribution always exists if the matrix $\M_{j,k} = t_{j+k}$ is positive semi-definite. 
A word of caution: one should not confuse the probability distribution entering in the Hamburger problem with the measure over the random matrices. 
The probability distribution relevant for the Hamburger problem is the eigenvalue density. 
At infinite $N$, the eigenvalue density is a deterministic variable (it does not fluctuate); the random variable whose moments we are computing is an eigenvalue chosen at random from the large matrix.


There are various generalizations of this problem. The {\it truncated moment problem} asks for necessary and sufficient conditions when only a subset of $T$ is given \cite{curto2000truncated}. This is relevant for the practical problem at hand, where we only compute a subset of correlators.
In the multi-matrix case, the analog of $t_k$ are traces of arbitrary ``words'' modulo cyclicity. So positivity in the multi-matrix case can be viewed as a non-commutative, multivariable generalization of the moment problem, see \cite{burgdorf2012truncated}.

\subsection{Multi-matrix models and the general algorithm}
For multi-matrix models, the space of correlators is exponentially bigger; we need to consider not just powers of $M$ but ``words'', e.g. $\phi = A + B + AB + \cdots + AB^2ABAB + \cdots$. Note that a generic off-diagonal element of $\mathcal{M}$, such as $\ev{ \tr A^2 B^2 A B}$ can be complex but $\mathcal{M}$ will always be Hermitian. If the model has a transpose symmetry, then $\M$ will be real and symmetric.


Let us now spell out the matrix bootstrap in generality. We start with the large $N$ loop equations, which are a set of quadratic equations in the single-trace correlators $\M$. There are infinitely many such equations, indexed by $a$. 
We set aside a small subspace of correlators $S$, which we call the search space. We choose this space so that if $S$ is determined, the loop equations will determine the rest of the correlators. For each point in $S$, we compute as many correlators as possible, assemble them into the inner product matrix $\M$, and check to see whether $\M \succeq 0$. (Even without the loop equations, there may be some positivity requirements on $S$; for example, if $S$ is the space $(t_1, t_2)$ we should impose $t_2 \ge t_1^2$.) The region of $S$ where this constraint is satisfied is our ``allowed region.'' Note that for some choices of $S$, the loop equations may not uniquely determine other correlators. For example, there may be a branch cut leading to multiple solutions. In such cases, the allowed region of $S$ consists of points where at least one solution has a positive $\M$.

In practice, it is of course impossible to compute infinitely many correlators. If one considers a large number of correlators, it may also be difficult to repeatedly compute eigenvalues and check for positivity. One can imagine a variety of approaches, where only some of the correlators are computed, or only a subset of the constraints are checked. How to achieve the best performance with limited computational resources is of course an important engineering problem. 

In general, one hopes that the allowed region converges to the exact solutions as one increases the number of constraints. Given a finite subregion, one estimate of the solution (assuming it is unique) is to maximize over $S$ the smallest eigenvalue of $\M$. 
For some applications, one is less interested in finding the allowed region; instead one wishes to simply know whether the allowed region is empty or not. This tests whether the model is self-consistent. In such a case, one can use, e.g., gradient ascent on the smallest eigenvalue of $\M$ and stop once the eigenvalue becomes positive.

\section{Bootstrapping 1-matrix models\la{bootS}}

\subsection{Single Hermitian matrix}
For simplicity, we start with the single Hermitian matrix model
\eqn{V(M) = \hf  M^2 + {g \over 4} M^4.}
We will first consider the case $ g>0$. We will treat the case $ g < 0$ and $g>0$ separately to emphasize some of the special features of this model. 
For the convenience of the reader, the exact solution of this model with our chosen conventions is reviewed in Appendix A.
We will take the search space $S$ to be a single parameter $t_2 \ge 0$ and set all odd correlation functions to zero. 

We follow the general approach outlined above to derive constraints. Starting from some value of $t_2$, we use the loop equations to compute all correlation functions up to some power $2d$. We assemble these correlators into the inner product matrix $\M_{d\times d}$, and find the region where all its eigenvalues are positive.

\begin{figure}[H]
\begin{center}
\includegraphics[scale=.7]{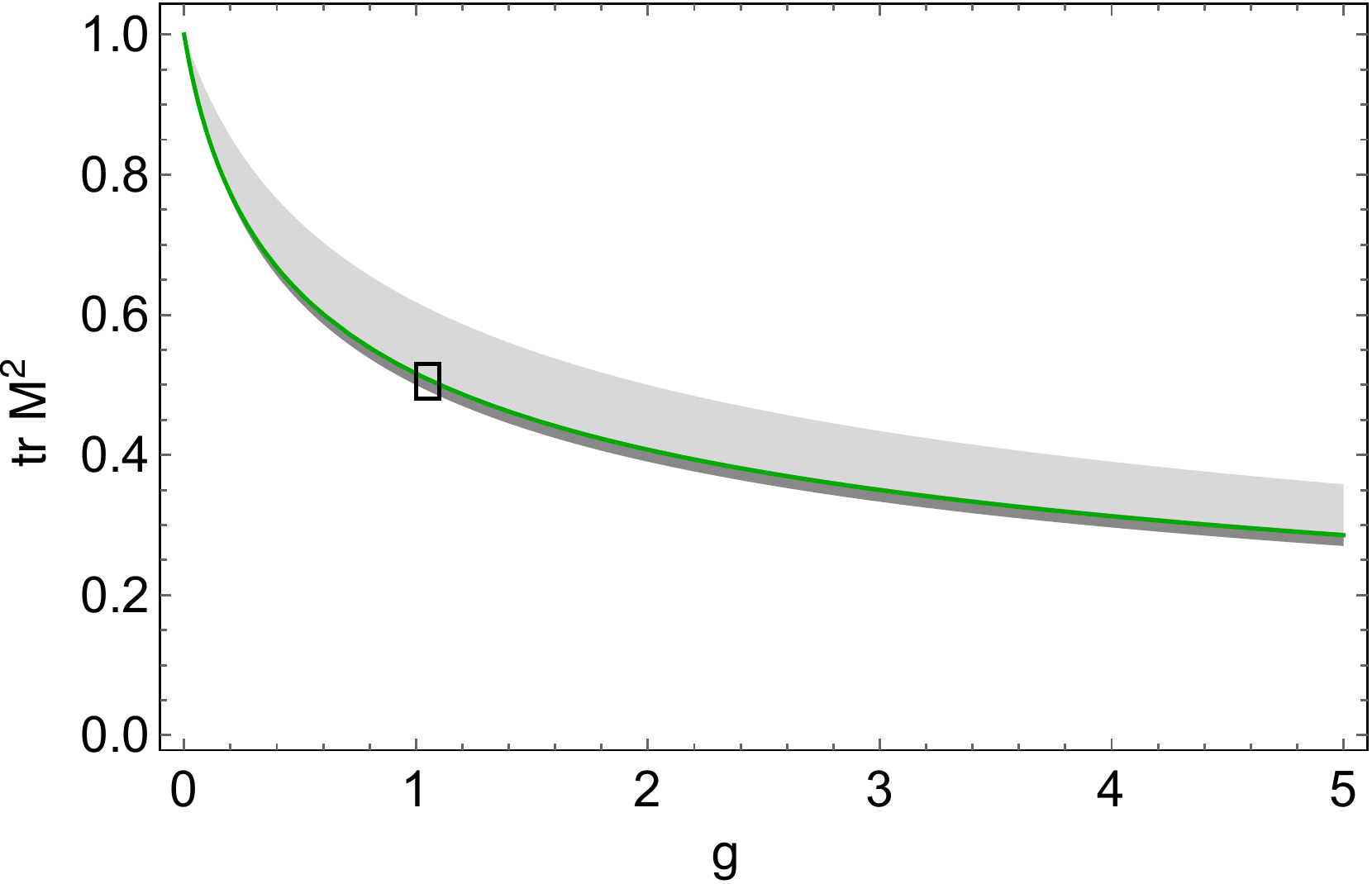}
\includegraphics[scale=.7]{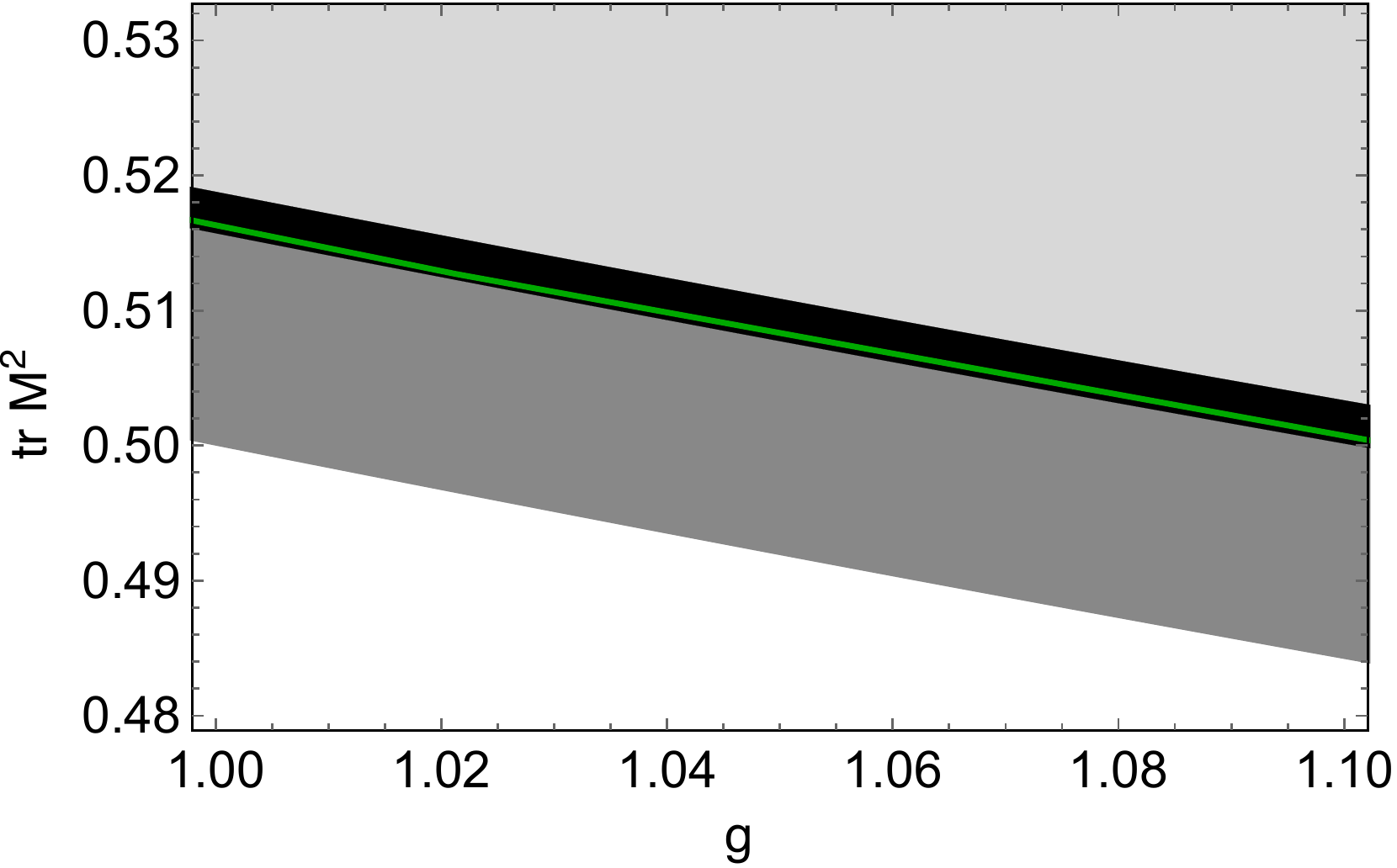}
\caption{The correlation function $t_2 = \ev{\tr M^2}$ as a function of the coupling $g$. Here we plot constraints from $\M_{d\times d} \succeq 0$, where the entries of $\M_{d \times d}$ are correlation functions up to $t_{2d} = \ev{\tr M^{2d}}$. The two shades of gray come from $d=4$ and $d=5$. The constraints from $d=6$ are so tight that they are indistinguishable from the solid green line (the exact solution) in the upper panel. In the lower panel, we show the constraints zoomed in on the small rectangle displayed in the upper panel, with the $d=6$ constraints in black. Note that for $g>1/12$ we are outside the radius of convergence of perturbation theory, so the bootstrap approach clearly does much better than a naive perturbative calculation of the same order $\sim 2d$.  }
\label{t2}
\end{center}
\end{figure}

From figure \ref{t2}, it is clear that the bootstrap approach converges rapidly to the exact solution. Furthermore, once bounds on $t_2$ is known, one can easily calculate bounds on any correlators $t_k$ by using the loop equations. For example, $t_4 = \frac{1-t_2}{g}$.

\begin{figure}[H]
\begin{center}
\begin{tabular}{c}
\includegraphics[scale=.7, trim = -73 28 0 0]{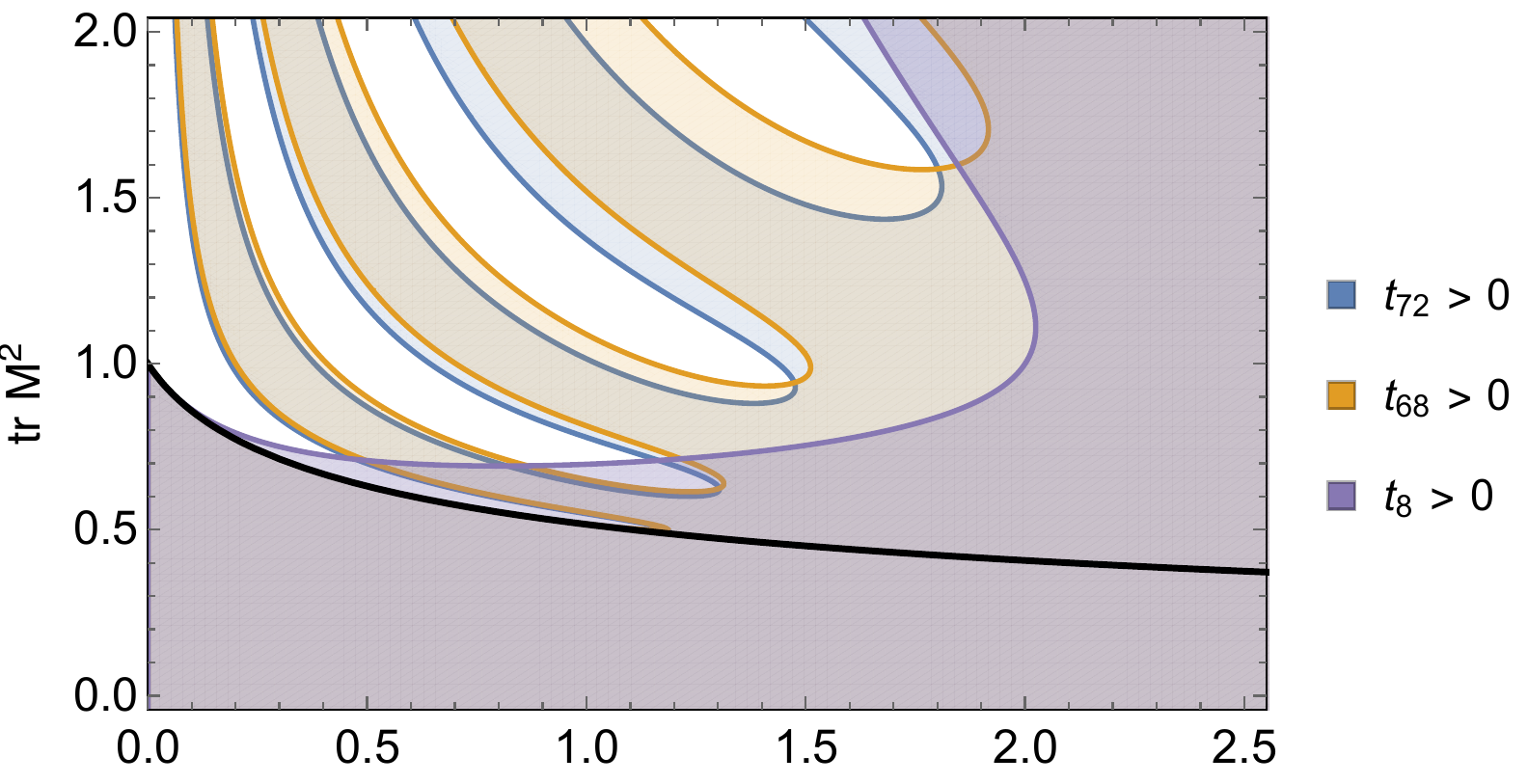}\\[20pt]
\includegraphics[scale=.7, trim = 0 0 0 0]{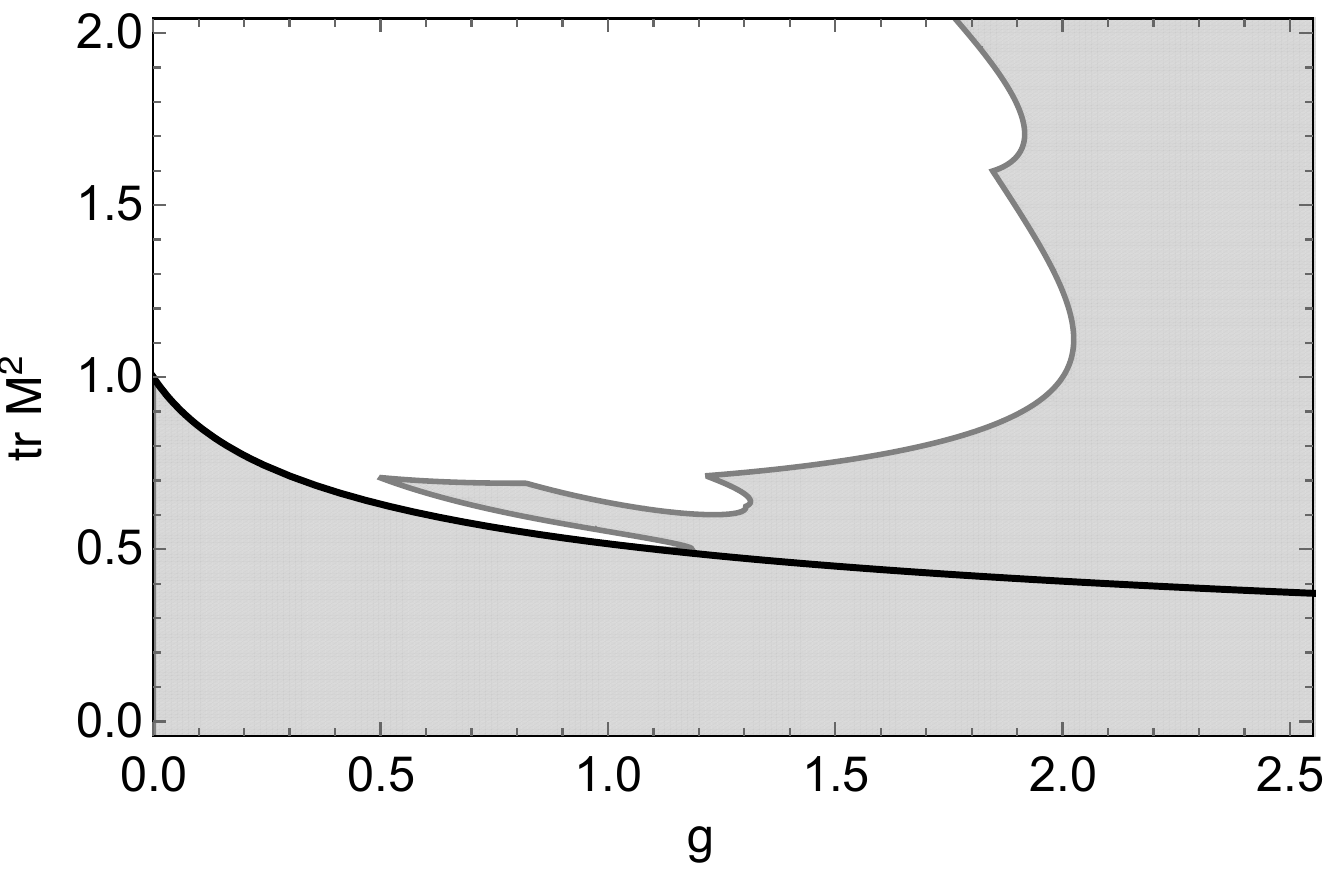}
\end{tabular}
\end{center}
\caption{Constraints from $t_k >0$, for $k\in \{ 8, 68, 72\} $. In the lower panel, we show the allowed region once all three constraints are imposed. The exact solution is indicated by the solid black curve. }
\label{fingers}

\end{figure}

It is interesting to try to ``look under the hood'' of the approach. One can do this by looking at the constraints coming from, e.g., single correlators. It follows from the loop equations that a correlator $t_{2k}$ will be a polynomial in $t_2$. The polynomial will typically have many zeros on the real axis, increasing with the degree of the polynomial. The location and number of zeros will depend on $g$. This is displayed in figure \ref{fingers}. If one considers the constraints from multiple correlators, the overall allowed region will be the intersection of the individual regions.


In general, if we plot the constraints coming from positivity of even correlators up to a certain fixed degree, we will always get a larger region than if we were to use the full positivity of $\M$. In practice, the difference is substantial, the convergence of the region allowed by positivity of the full inner product $\M$ is much faster.

One might find it surprising that the bootstrap method even converges at all. Why is it that positivity is so strong that only the correct solution is allowed, as opposed to, e.g., some finite island? Actually, convergence of this method would be naturally explained if the {\it exact} solution has a null vector, e.g., if we can find some matrix $\phi$ such that $\tr \phi^* \phi = 0$. If such a vector existed, then a small perturbation of the matrix $\M$ could easily violate $\M \succeq 0$. Geometrically, the constraint $\M \succeq 0$ means that the allowed region is a cone; a (nearly) null vector would mean that the exact solution lies (nearly) on the boundary of the cone. 

This criterion might seem exotic, but in fact it is satisfied by all models where at least one of the matrices in the ensemble has an eigenvalue distribution which has support on finite interval(s). Then if we are allowed to consider polynomials in $\lambda$ with large degree, we can approximate a function which is zero on the support of the eigenvalue distribution but non-zero elsewhere. Furthermore, as we increase the degree of the polynomial, we expect to be able to better and better approximate such a function. This would then naturally explain the convergence of the method.

\begin{figure}[H]
\begin{center}
\includegraphics[scale=.65]{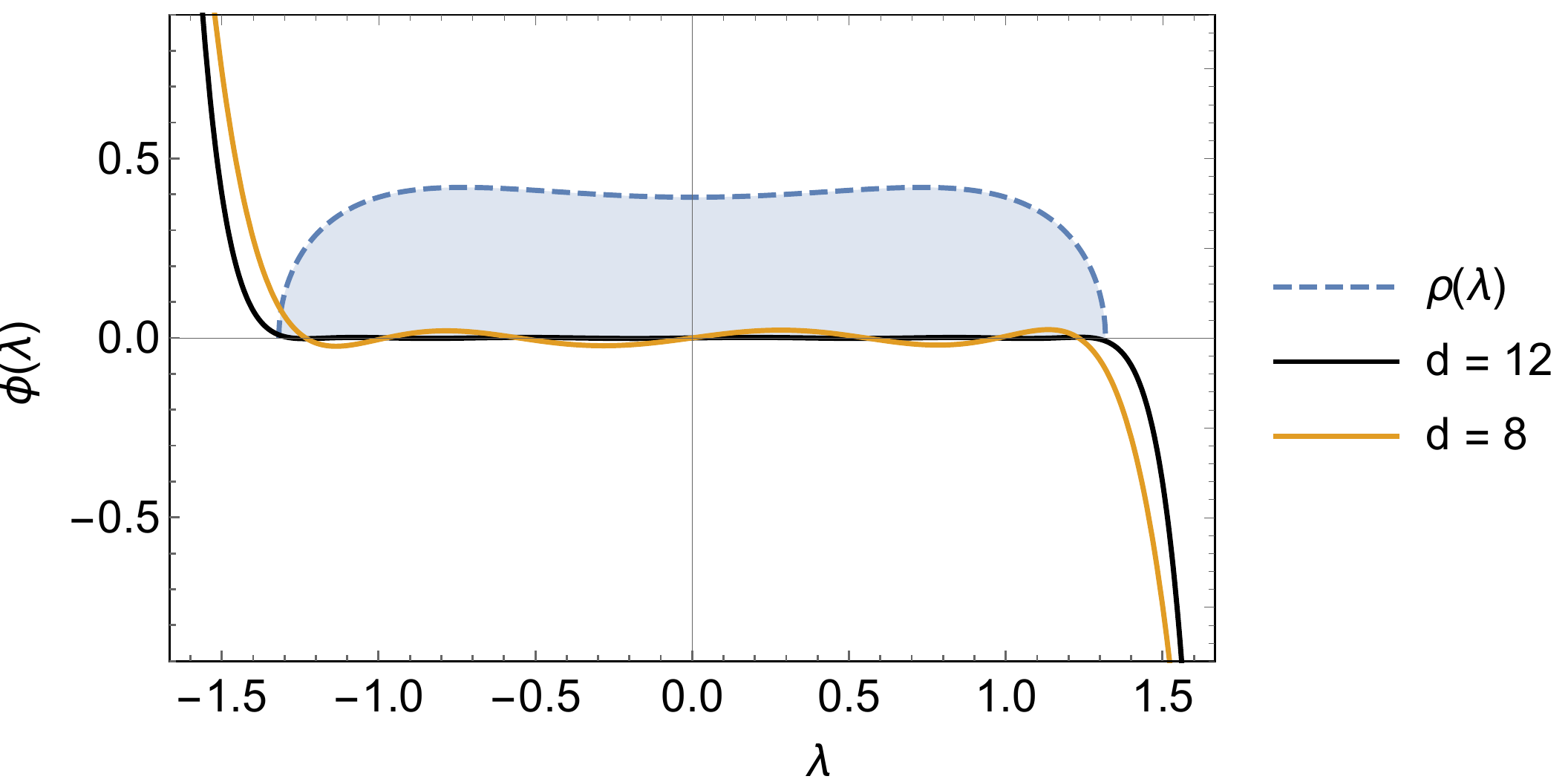}
\caption{Nearly null eigenvectors of $\M$ for the exact solution with $g=1$. We plot in shaded blue the eigenvalue distribution $\rho(\lambda)$, which has compact support. The solid curves are the polynomials in $M$ (or equivalently in $\lambda$) which correspond to eigenvectors with lowest eigenvalue of $\M_{d \times d}$ for $d=8$ and $d=12$. As the degree increases, the eigenvectors nearly vanish in the region where the eigenvalue distribution $\rho(\lambda)$ has support. As $d\to \infty$, we expect that there are many  nearly null eigenvectors of $\M$. This explains why the bootstrap approach works well in the single matrix case. (Note: please do not confuse $\lambda$, the eigenvalues of the random matrix $M$ with the eigenvalues of the matrix of correlators $\M$.) }
\label{null}
\end{center}
\end{figure}

We can test this explanation by simply plotting some eigenvectors of $\M$ in the exact solution which have small eigenvalues. This is done in figure \ref{null}.
Notice also that this explanation also predicts (correctly) that convergence will be much improved when we use positivity of the full matrix $\M$ as opposed to just positivity of even correlators $t_{2k}$, since the tightest constraints come not from monomials but from the special polynomials which nearly vanish on the support of $\rho(\lambda)$. 

In the single matrix model, the eigenvalue density is supported on some finite interval(s). However, it seems reasonable that the method will perform well even when this condition is not exactly true. For example, if the eigenvalue distribution of some matrix $M$ rapidly decays faster than any power of $\lambda$ outside some interval, there should be many high degree polynomials in $M$ which have nearly zero norm. In a multi-matrix model, $M$ could be any composite matrix, built out of powers of the matrices that are integrated over. If the action $S$ is a polynomial in $A_1, \cdots A_k$ that is bounded from below, it seems likely that this condition will hold. 
For some special potentials, it might be possible to argue that the eigenvalues of a matrix continue to have bounded support; for example, in a potential like $V= \hf A^2 + \hf B^2 + g (A+B)^4$. For $g=0$ the eigenvalues of $A$ are distributed like a semi-circle and hence bounded. As we turn on $g > 0$, the interaction should provide an additional confining force for $A$. So we expect the eigenvalues of $A$ to remain bounded.


\subsubsection{Unbounded potentials and the tip of the peninsula}
An interesting test of our method is to go to negative values of the coupling $-1/12 < g <0$. One motivation is that the limit $g \to -1/12$ is the physically interesting regime to make contact with string theory, since the typical number of interaction vertices (interpreted as the area of the planar diagram) is becoming large.

Another motivation is that in this regime, the matrix ensemble is not well-defined for finite $N$ since the potential is unbounded from below, so a Monte Carlo simulation would be problematic (or at least subtle). This subtlety does not arise in the bootstrap approach, which works directly at large $N$. Nevertheless, we see an interesting behavior in the constraints for negative values of $g$. Applying the general method discussed above, the allowed region converges rapidly to the exact solution. Using correlators up to degree 20, the width of the allowed region is $\lesssim 1\%$ in $t_2$ in the negative region. 

\begin{figure}[ht]
\begin{center}
\includegraphics[scale=.75]{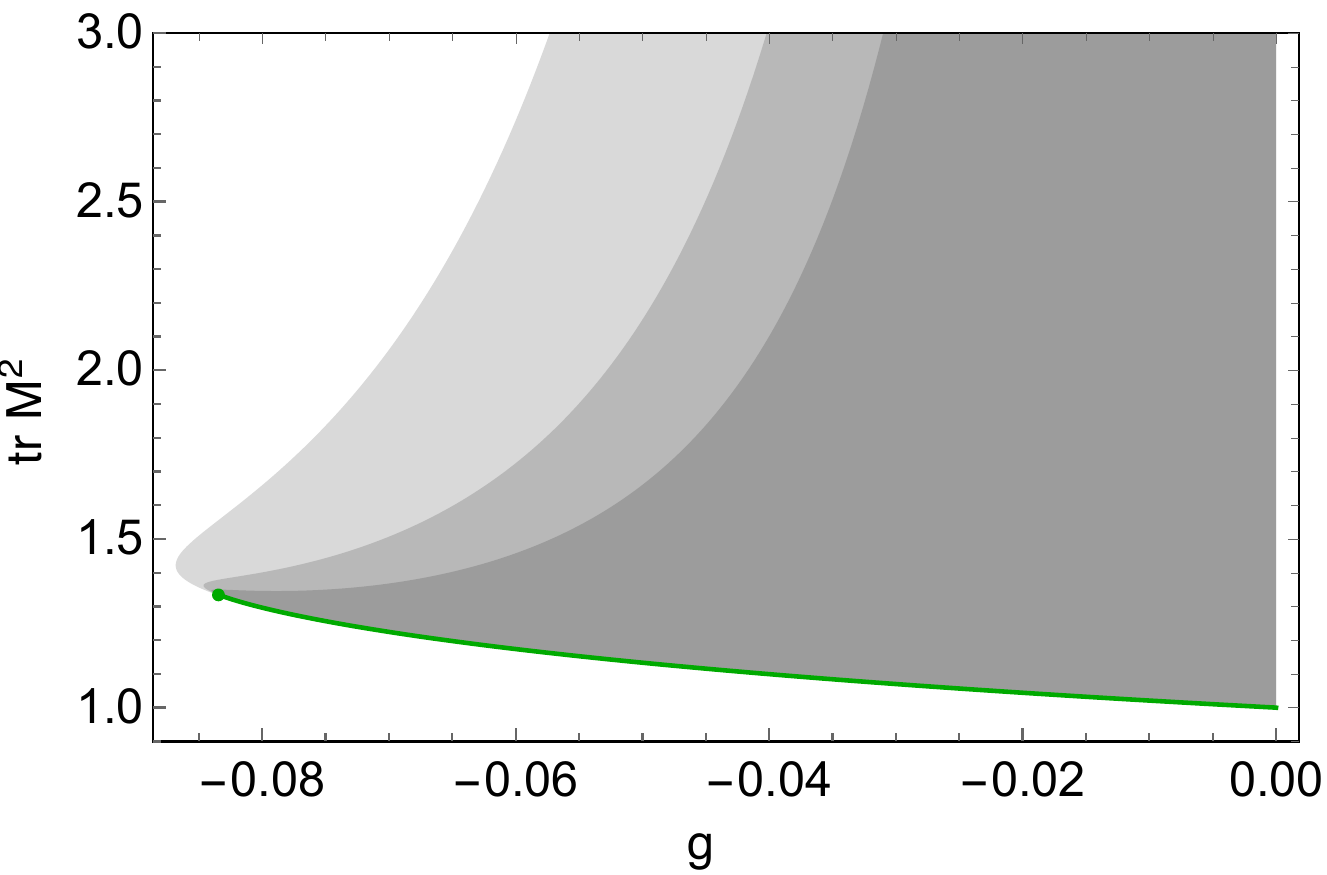}
\caption{A view of the ``peninsula''. Here we consider constraints coming from $t_k > 0$. For this plot, we consider $k=32,52,72$. This leads to the shaded gray region. As more constraints are enforced (higher $k$), the peninsula is eroded down to a smaller region. The green curve is the exact solution, which is close to the boundary of the allowed region. At large $k$, the tip of the peninsula approaches the critical point of the exact solution $g_* = -1/12 \approx -0.083$. This demonstrates that the numerical method can be used for finding the critical point of the model.
If we use positivity from the full matrix $\M \succeq 0$, the constraints are indistinguishable from the green line at $d \sim 20$.
}
\label{peninsula}
\end{center}
\end{figure}

To get a clear picture of what is going on, it is again instructive to consider constraints coming from positivity of single even-degree correlators. 
This is shown in figure \ref{peninsula}. An interesting feature is that the constrained region looks like a ``peninsula.'' Beyond the tip of the peninsula, no value of $t_2$ is allowed. This means that such models are inconsistent with positivity. In fact, if we look at the exact solution, beyond the critical value $g=-1/12$, the correlators do not exist; formally, the value of $t_2$ becomes complex. Note that the tip of the peninsula happens when there is a double zero. If we are computing the constraints from $\M$, the tip occurs when the smallest eigenvalue $m_1$ of $\M$ satisfies $m_1(t_2) = \pd m_1/\pd t_2 = 0$.



One might wonder why the exact solution in figure \ref{peninsula} is quite close to the bottom of the peninsula. For the exact solution to approach the lower bound, we must be able to neglect $t_k$ for large $k$ in the loop equations. 
This in turn is equivalent to neglecting high powers of $g$. But this is equivalent to truncating perturbation theory at a finite order. Since we are within the radius of convergence of perturbation theory, this is not too surprising. 

A somewhat different perspective on the $g<0$ computation is the following. One might forget about how we derived the loop equations, and simply view them as a set of rules for computing correlation functions of matrices. These rules are seemingly well-defined for any value of $g$. However, not all possible rules for computing correlation functions are sensible, as some will lead to violations of the positivity requirement \nref{masterpos}. Here we have demonstrated that the rules are not sensible beyond a critical value of $g_*$. This is similar to the CFT bootstrap philosophy: not all possible sets of dimensions and OPE coefficients are sensible.

An interesting question is the behavior of the free energy near the critical point. The critical exponent $\gamma$ defined by $F_\text{sing} \sim (g_c-g)^{2-\gamma}$ can be compared with a continuum methods (a string worldsheet calculation). 
To compute the free energy $F(g)$, one can integrate $t_4$ 
\eqn{F(g) = - {1 \over N^2} \log Z(g) = 4 \int t_4 \, dg .}
In this way, we can (in principle) extract from the bootstrap the critical exponent $\gamma$. Note that $\gamma$ is defined by the leading non-integer exponent in $(g_c-g)$. So it is convenient to compute derivatives of $F(g)$ and look for the smallest power $p$ such that $\pd_g^{p+1} F \propto \pd^p t_4$ is diverging. Derivatives of $F$ are related to connected components of multi-trace correlators. In practice, it may be easier to compute these than to estimate derivatives of $t_k$. A direct computation of the connected components involves going to higher order in $1/N$, see Appendix \ref{1overN}.

In general, the critical surface in the space of the couplings where the model ceases to be well-defined is the first step in identifying the infrared/continuum theory. 
In the single matrix model, and for very special multi-matrix models, one can completely characterize the continuum theory as a 2D minimal model coupled to Liouville gravity. By considering more general multi-matrix models, one might hope to extract more general string worldsheet theories.


\subsubsection{Other single-matrix models}
\begin{figure}[H]
\begin{center}
\includegraphics[scale=.75]{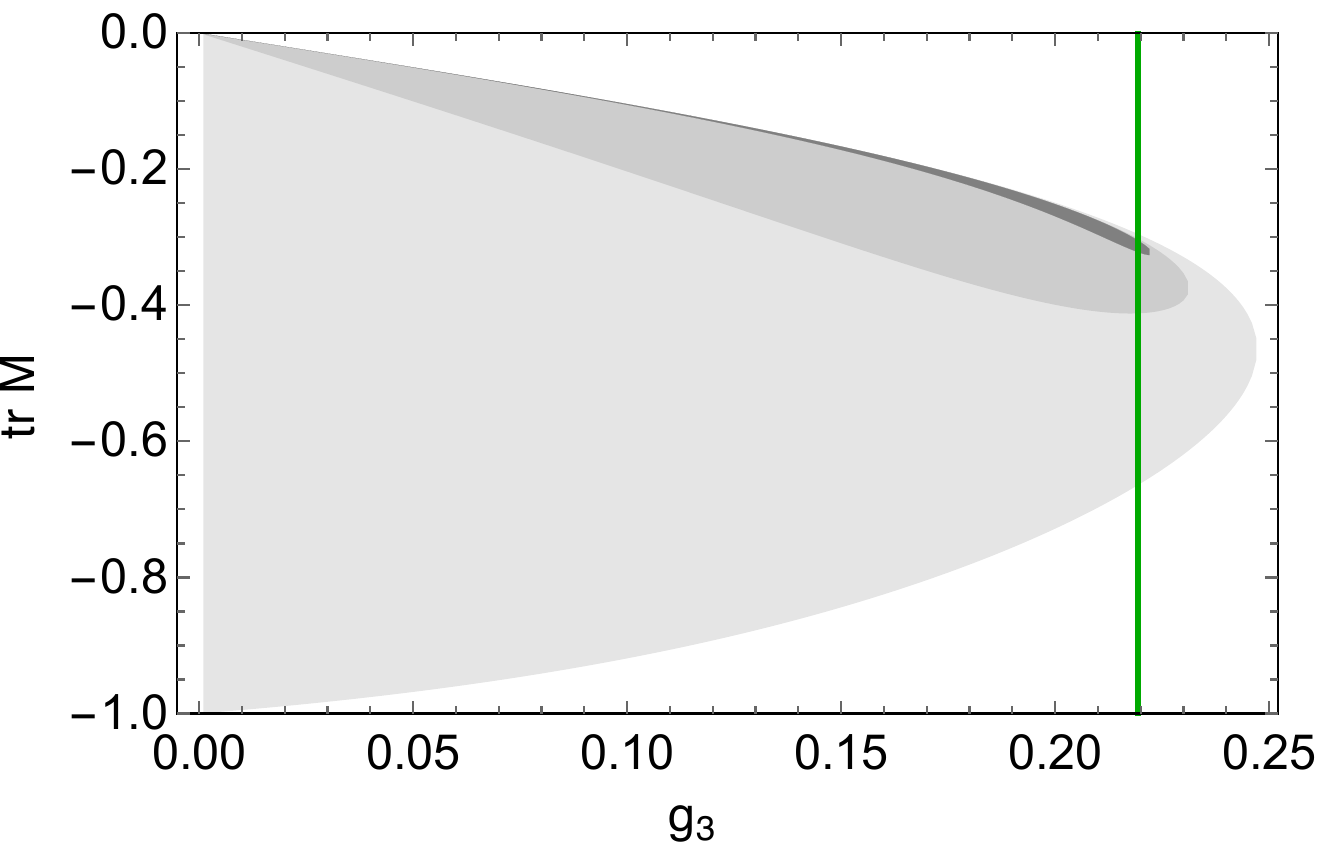}
\caption{The correlation function $t_1 $ as a function of the coupling $g_3$. Here we plot constraints from $\M_{d\times d} \succeq 0$. The concentric regions are from $d=3, 4, 6$. The solid green line indicates the critical coupling $g_3^2 = 1/(12 \sqrt{3})$. The constraints are symmetric under $M \to -M, g_3 \to -g_3$. }
\label{cubic}
\end{center}
\end{figure}

In addition to the quartic model, we also considered the cubic model:
\eqn{V(M) = \hf M^2 + {g_3 \over 3} M^3 }
No modification of the above method is needed to solve this model, except that the search space is the correlator $t_1$. Unlike the quartic model, the cubic model does not have a strong coupling region, since the model only makes sense on a finite interval in parameter space. From the exact solution, the critical values of the coupling are $g_3^2 \le 1/(12 \sqrt{3})$, which is again nicely reproduced by the bootstrap method, see figure \ref{cubic}.

\begin{figure}[ht]
\begin{center}
\includegraphics[scale=.55]{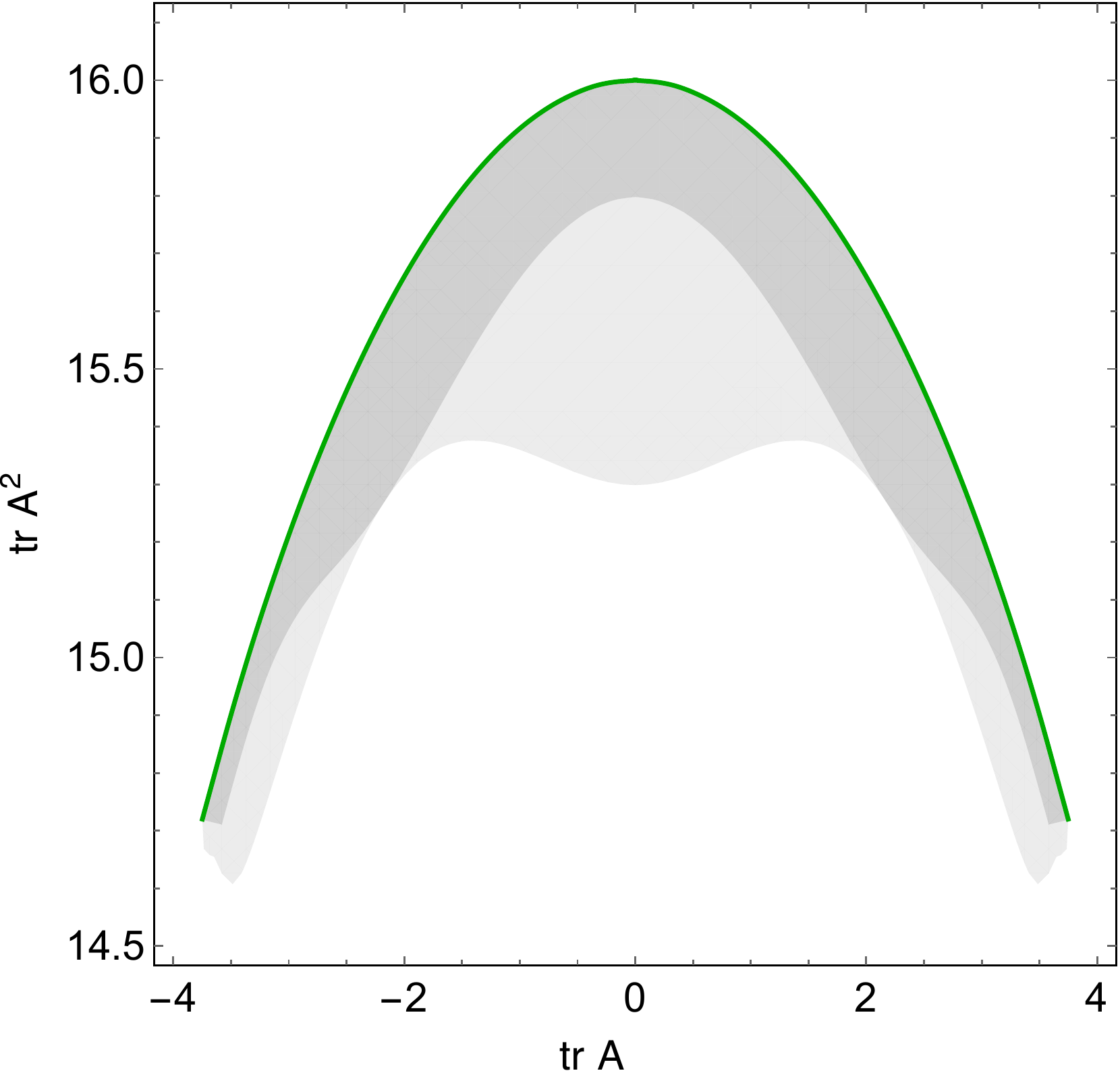}
\caption{Constraints from $\M_{d \times d} \succeq 0$ for the inverted potential $V = -\hf M^2 + {g \over 4} M^4$ for $g=1/16$. The allowed regions in gray are from $d=8$ and $d=9$. We also plot in green the 1-parameter family of exact solutions. These correspond to 2-cut solutions; the boundary of the green curve corresponds to a completely asymmetric 1-cut solution. The green curve is very close (but not exactly coincident with) the boundary of the allowed region. By going to higher degree $d=35$, we find good convergence to the green line to an accuracy of $\sim 0.01\%$. We also indicate the {\it a priori} constraint $t_2 \ge t_1^2$.
}
\label{symbreak}
\end{center}
\end{figure}

\begin{figure}[H]
\begin{center}
\includegraphics[scale=.6]{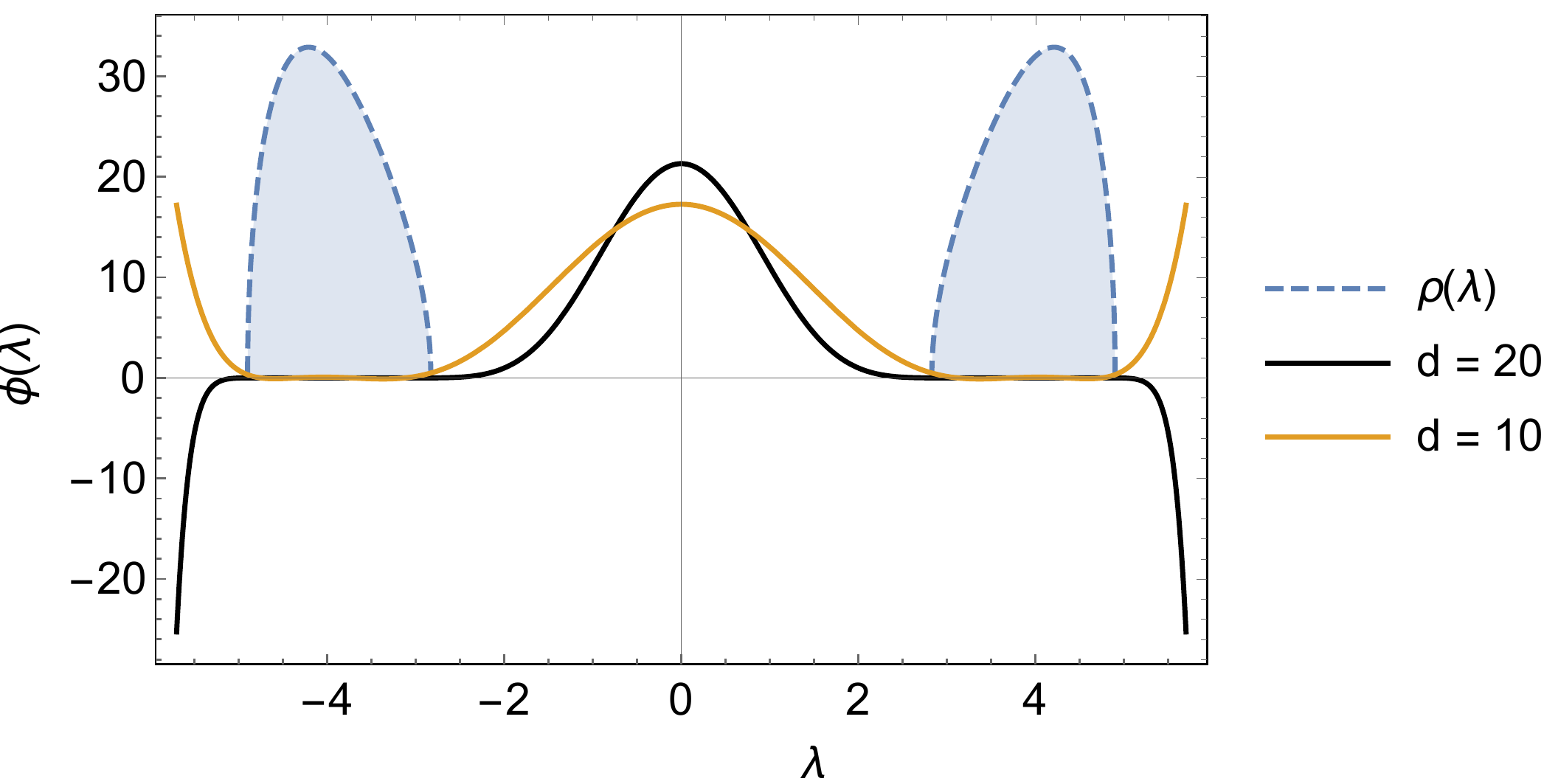}
\caption{Nearly null eigenvectors of $\M$ for the exact solution with $g=-1/16$ and $t_1 = 0$. We plot in shaded blue the 2-cut eigenvalue distribution $\rho(\lambda)$. The solid curves are the smallest-eigenvalue eigenvectors of $M$ for $d=10, 20$. As in figure \ref{null}, the eigenvectors nearly vanish in the region where the eigenvalue distribution $\rho(\lambda)$ has support. The normalization of the eigenvectors in this figure is arbitrary.}
\label{null2}
\end{center}
\end{figure}

We have so far considered a case where there is a unique classical solution in the large $N$ limit. In general, however, there could be a family of solutions. For a single matrix model, if the potential $V$ has multiple minima, there are solutions where a fraction of the eigenvalues sit near each minimum. (There is no tunneling of the eigenvalues at large $N$). We will consider the simplest possibility, where $V(M) = -\hf M^2 + {g \over 4} M^4$. We can read off the loop equations from \nref{loopeqn}, or we can derive them diagrammatically using the fact that the free propagator comes with a negative sign. 

In this model\footnote{This model has an interesting critical point whose string interpretation is described in \cite{Klebanov:2003wg}. The critical point occurs when the 2 cuts merge into a single cut.}, adjusting the filling fraction is equivalent to turning on $\ev{\tr M} = t_1$. So our search space is the two dimensional space $(t_1,t_2)$ subject to $t_2 \ge t_1^2$.  
A non-zero expectation value $t_1$ means that we are considering a case where the $Z_2$ symmetry $M \to - M$ is spontaneously broken. So we expect the method to converge to a curve in the space of correlators, as opposed to a point. This is illustrated in figure \ref{symbreak}. By increasing the degree of the correlators considered, we find good numerical evidence of a convergence to a 1 dimensional curve.

One could also consider matrix integrals with a potential involving double trace or higher interactions, with the coupling constants appropriately scaled with $N$ so as to maintain the 't Hooft planar limit. We expect the bootstrap approach to work well in this case as well.

\section{Bootstrapping multi-matrix models \la{bootMS}}
We now graduate from single matrices to multi-matrices. In the single matrix case, all polynomial potentials are solvable; this is not true for the vast majority of multi-matrix potentials. A notable exception is the Ising matrix model and its variants, which we discuss in appendix \ref{app:ising}.
In the Ising model, the only interaction between matrices is the quadratic interaction $\tr A B$. 
The next simplest case is the cubic interaction
\eqn{V=  W(A) + W(B) + h (A B^2 + B^2 A). \la{multi} }
This model has the same $\mathbb{Z}_2 \times \mathbb{Z}_2$ symmetry of the Ising model. One $ \mathbb{Z}_2$ is generated by $A \leftrightarrow B$; the other is generated by $A \to A^T, B \to B^T$. The latter guarantees that all correlators $\ev{ \tr A^j B^k A^l B^m \cdots}$ are real.
To our knowledge, the above model is not solvable for a generic potential $W$. For a 2-matrix interaction, an interaction of the form $A^n B^m$ can be reduced to an integral over 2 eigenvalue densities using the Itzykson-Zuber formula \cite{Itzykson:1979fi}, but this trick does not extend to an interaction of the form $h(A^n B^m + B^m A^n)$. So this model provides a proof-of-concept that the bootstrap gives highly non-trivial constraints on a model that would be difficult or impossible to analyze analytically. 

An exception is when $W$ is a cubic polynomial, in which case the model is solvable. We discuss this in \ref{app:cubic}; we found it was nonetheless a useful exercise to solve the cubic potential using the bootstrap approach, since the dimension of the search space is quite small, $\dim S = 2$.
\def\fminsdp{\texttt{fminsdp}}
To demonstrate that the method works on a model that is not believed to be solvable, we will take the above interaction with $W(A) = { 1\over 4} A^4$ and $h=1$. This model has a higher dimensional search space. For correlators up to degree 10, we found that $\dim S =8$.
To obtain constraints, we use the Matlab package \href{https://www.mathworks.com/matlabcentral/fileexchange/43643-fminsdp}{\fminsdp} which tests for matrix positivity by attempting a Cholesky decomposition of the inner product matrix $\M = {\cal L} {\cal L}^\dagger$ where $\cal L$ is a lower triangular matrix with positive real entries on the diagonal. One can find the allowed region in search space by minimizing a loss function $L$ subject to the positive semi-definite constraint on $\M$. To derive constraints on a subspace of the search space, e.g., $(t_1, t_2)$, one can choose $L = (\cos \theta) t_1 + (\sin \theta) t_2$ and sweep through $\theta$. The resulting output will be an 8 dimensional curve parameterized by $\theta$. Projecting it into the $(t_1, t_2)$ plane gives the allowed region.

\begin{figure}[ht]
\begin{center}
\includegraphics[scale=.75]{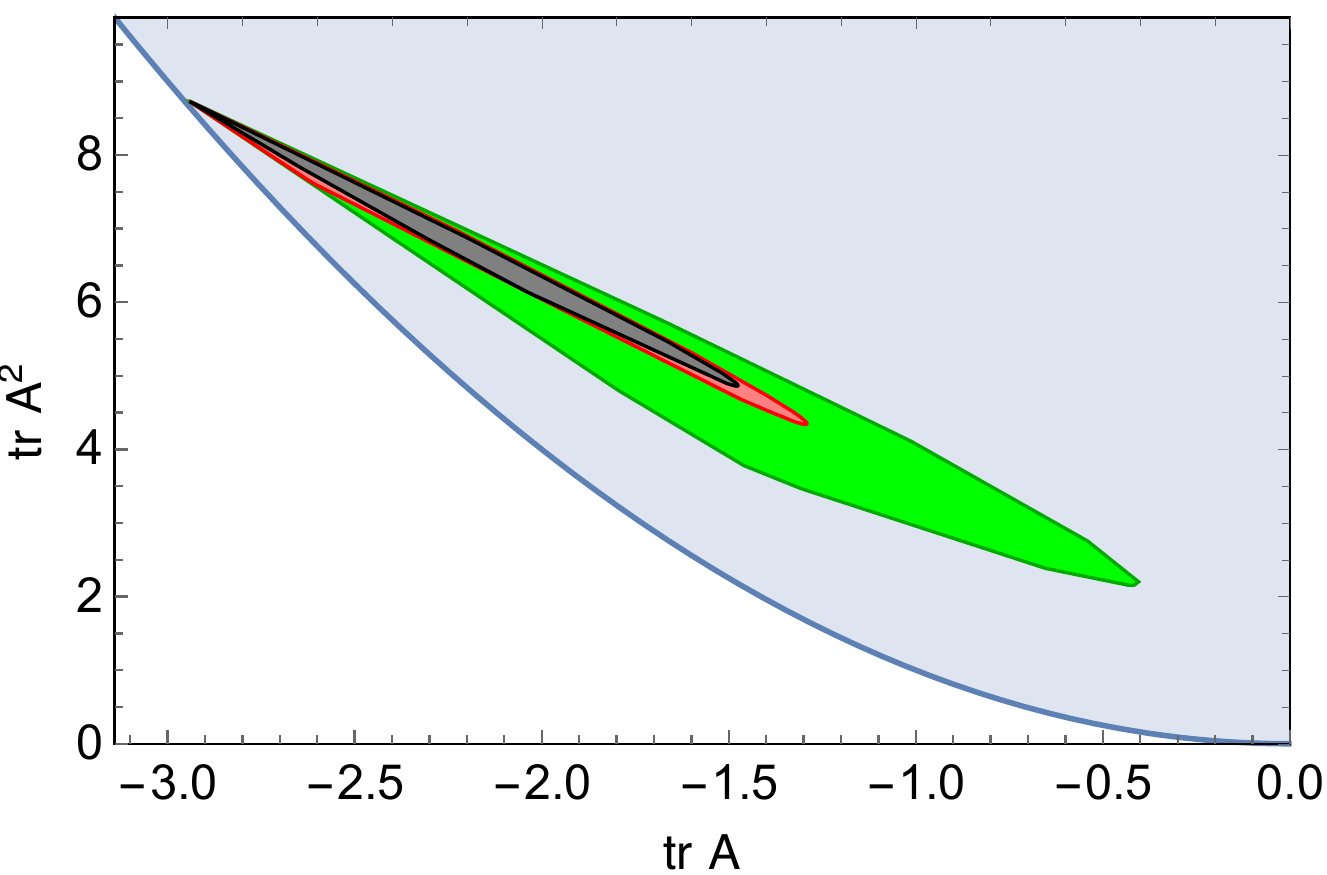}
\caption{Constraints for the quartic model with an inner product matrix of size $k = 15, 25, 35$.  The shaded blue region $t_2 \ge t_1^2$ is the {\it a priori} constraint on the search space. }
\label{hard}
\end{center}
\end{figure}

In practice \fminsdp \: works well if a feasible initial point is provided. One can obtain such an initial point by simply performing gradient ascent on the minimum eigenvalue of $\M$ and stopping once the eigenvalue is positive. We plot the resulting bounds in \ref{hard}. One can estimate the numerical uncertainty of the contours by varying the initial starting point and comparing the resulting contours. We found stability to within a few percent.


We organized the inner product matrix by increasing degree of correlators in lexicographic order. By consider larger and larger matrices, we found some mild evidence of convergence to a point in correlator space, see figure \ref{hard} where constraints up to $k = 35$ are shown. By increasing the size of the constraint matrix to $k=45$, we checked that the points on the boundary are ruled out.
There are clearly many more questions that could be asked about this particular model. For example, can we characterize the critical points of this model for a more general potential $W(A)$ and arbitrary values of $h$? We hope to return to these questions in future work.

\section{Discussion \la{disS}} 
In this paper, we proposed a new method to solve multi-matrix models. 
We showed that the method works extremely well on integrable models and explained why it works in this case. We focused on simple models in part to gain confidence that the bootstrap method works, since the exact solutions of these models is known. 
We also applied the method to a 2-matrix model that is not known to be solvable and extracted relatively tight constraints.

One feature of the matrix bootstrap is that it is computationally efficient.
Most of the computations in this paper could be done with only a few seconds of CPU time. The exception is the 2-matrix model considered in section \ref{bootMS}, where a higher dimensional optimization problem had to be solved. In all of our plots, we saw evidence of convergence just by considering relatively small inner product matrices with just a few dozen rows and columns.
No doubt with more computational resources (and more efficient algorithms), we could obtain extremely strong constraints from higher correlators; the main point is that even with extremely modest computational resources one can derive constraints that would be hard to obtain by any other method.

Besides the obvious goal of solving more multi-matrix models involving higher degree polynomials and/or more matrices, there are a variety of future directions. We list a variety of other complications that should be addressed in the future:
\begin{enumerate}
	\item As we have already discussed, it would be exciting to find new critical points using this approach. In the past, matrix descriptions of ``minimal string theories'' were found, e.g., minimal model CFTs coupled to Liouville gravity. It would be interesting to find alternative matrix descriptions with the same continuum description, or even better, completely new continuum models. The usual conformal bootstrap can be used to (re)-discover CFTs; perhaps the matrix bootstrap can be used to find (in general non-unitary) 2D CFTs coupled to gravity!
	\item Can this method be used to solve models with fermions? Some supersymmetric matrix models are of interest, for example, the IKKT model \cite{ikkt}. If the theory is quadratic in the fermions, one can integrate them out and obtain a determinant. If the determinant is not a positive function of the bosonic fields, we cannot run the same arguments based on positive measures. In a similar vein, we could ask whether this method could be extended to solve models with complex couplings.
			
	\item We have focused on zero-dimensional matrix models. One could wonder whether this method could be used to solve higher-dimensional models. If we go up just one dimension (Euclidean matrix quantum mechanics), we could ask whether, e.g., the $c=1$ matrix model could be solved by this method (see \cite{Klebanov:1991qa} for a review). If we discretize Euclidean time, we are left with a model that is a matrix chain. We believe that such a model could likely be solved by our method; perhaps it is possible to modify the approach so that it works directly in the continuum.
	\item Does this method work for other large $N$ theories, e.g., tensor models \cite{Klebanov:2018fzb}? 
	\item In Appendix \ref{1overN} we discuss the possibility of computing $1/N$ corrections using a similar approach. Does this work in practice? Is it possible to do better, e.g., go to the double scaling limit?

\end{enumerate}
A dream would be that some sort of bootstrap method could be used to numerically solve any matrix theory, such as large $N$ QCD, in the 't Hooft limit \cite{Anderson:2016rcw}. This seems like a far-fetched dream, but one worth having nonetheless.

\section*{Acknowledgements}
I thank Adam Brown, Igor Klebanov, Sam Ritchie, Daniel Roberts, Steve Shenker, and Douglas Stanford for helpful discussions and corrections. Particular thanks goes to Juan Maldacena for many stimulating conversations on matrices and beyond and for suggestions on the manuscript.
I am supported in part by an NDSEG fellowship.

\appendix

\section{Review of the single matrix model \la{exactA}}
Here we review the solution of the loop equations in the 1-matrix case. (See \cite{Eynard:2015aea} for a modern review). The strategy is to exchange the infinite number of real variables that appear in the loop equations for an unknown function. 
\eqn{\sum_k x^k \ev{\tr M^k V'(M)} = \sum_k \sum_{\ell=0}^{k-1} \ev{\tr M^{\ell}} \ev{\tr M^{k-\ell -1}}x^k . \la{sumx}}
It is useful to define the resolvent $R$ and a closely related function $P$:
\eqn{R(x) &=\ev{\tr {1 \over x-M} } = \sum_{k=0} t_k x^{-k-1}, \\
	P(x) &=  \ev{\tr {V'(x)-V'(M) \over x-M}} .
}
$P$ must be an analytic function, since the potential poles at the eigenvalues of $M$ are cancelled by the zeros in the numerator. Furthermore, at large $x$, $P \sim V'/x$, so if $V$ is a degree $d$ polynomial, $P$ must be degree $d-1$. 
In terms of these functions, the loop equation \nref{sumx} becomes a simple quadratic equation:
\eqn{V'(x) R(x) - P(x)= R(x)^2,}
the solution of which is
\eqn{R(x) = \hf \lp V'(x) \pm \sqrt{(V')^2 -4P}\rp.}
We may write the term in the square root $(V')^2 - 4 P = M^2 \sigma$, where $\sigma$ only has simple roots $M^2$ has roots with an even multiplicity.
\eqn{R(x) = \hf \lp V'(x) + M(x)\sqrt{\sigma(x)}\rp.}
So far, we have found a solution to the loop equations involving a polynomial of degree $d-1$. We have one more condition coming from $R(x) \sim 1/x$ at large $x$. So we obtain $d-2$ unknowns, which agrees with our estimate of the dimensionality of the search space.

Now the branch cut of the resolvent famously determines the eigenvalue density $\rho(\lambda)$, since
\eqn{R(\lambda) = \int d \mu {\rho(\mu) \over \lambda - \mu}. }

\subsection{Single cut}

In the simplest case, we assume that all the eigenvalues live on a single interval.
This motivates the following ansatz for $R$:
\eqn{R(\lambda) = \hf V'(\lambda) +  P(\lambda) \sqrt{(\lambda-a_1)(\lambda-a_2) }}
where $P$ is some analytic function. Now since $\rho$ integrates to 1, we must have $R(\lambda) = {1 \over \lambda} + \cdots $  in a large $\lambda$ expansion. This seemingly trivial condition determines the function $P$. 

Let us consider the case $V = \hf \lambda^2 + {g \over 4} \lambda^4$, following \cite{bipz}. We have $R = \hf \lp \lambda + g \lambda^3\rp + P(\lambda)\sqrt{\lambda^2 - a^2} $ where we assume that the eigenvalue distribution is symmetric  $a_2 = - a_1 = a$.
If $P$ contains terms that are higher order in $\lambda^2$, then there is no way that $R$ can have the right asymptotics. \
So $P$ must be quadratic in $\lambda$ and in fact there are enough conditions to determine $P = -{g \over 2}\lambda^2 - {1\over 4}ga^2 - \hf $ and the location of the branch cut satisfies $a^2  (4 + 3 g a^2 ) = 16 $. 
Now the discontinuity in $R$ around the branch cut gives
\eqn{\rho(\lambda) \propto \lp {g \lambda^2 } + {g a^2 \over 2} + 1  \rp\sqrt{\lambda^2 - a^2 } }
Note that there two solutions $a^2 = {2 \over 3g} \lp -1 \pm \sqrt{1+ 12 g } \rp$. For $g>0$, we must choose the $+$ sign so that $a^2$ is positive. For $g<0$ both solutions can have $a^2 > 0$. However, requiring positive eigenvalue density forces the $+$ sign solution. Notice from the exact solution the critical point $g_* = -1/12$. Beyond this point $a^2$ is complex.
\subsection{Multi-cut solutions}
Now we consider the inverted potential $V = - \hf \lambda^2 + {g \over 4} \lambda^4$. 
We first search for a single-cut solution. We find 
\eqn{P(\lambda) \propto -\hf g \lambda^2 - {g \over 4} \lp a_1 + a_2 \rp \lambda + {1 \over 16} \lp 8-3 a_1^2 g - 2 a_1 a_2 g - 3 a_2^2 g \rp }
We also get 2 conditions on $a_1$ and $a_2$:
\eqn{(a_1 + a_2) (5 a_1^2 g-2 a_2 a_1 g+5 a_2^2 g-8) =0, \\ \left(a_1-a_2\right){}^2 \left(3 \left(5 a_1^2+6 a_2 a_1+5 a_2^2\right) g-16\right)=256}
There is a positive value $g_* = 1/15$ such that symmetry is restored $a_1 = -a_2$.
Below this value of $g$ there is a 2-cut solution with support on $(a_1, a_2)$ and $(-a_1, -a_2)$. The additional undetermined parameter in this phase can be interpreted as the filling fraction $f$ between the two minima.
Here it is important to exclude the solutions where the eigenvalue density is not positive. We can also see that for $g>1/4$, there is a single-cut symmetric solution.

We also search for a 2-cut solution. This means we take the ansatz
\eqn{R(\lambda) = \hf V' + P \sqrt{(\lambda - a_1)(\lambda-a_2) ( \lambda - a_3) ( \lambda - a_4) } \la{2cut}}
where $P(\lambda)$ is a degree 1 polynomial. The solution to the constraint $R(\lambda) \sim 1/\lambda$ then determines $P(\lambda)$ uniquely in terms of $a_i$. Furthermore, we get 3 constraints on $a_i$, which means we have a 1-parameter family of solutions.
The simplest solution of this equation is the symmetric case, where $a_4 = -a_3$, $a_2 = -a_1$. This was analyzed in \cite{Cicuta:1986pu}. The solution exists when $g< 1/4$. This solution has very simple expressions for the correlators:
\eqn{t_1 = 0, \quad t_2 = 1/g.}
For the general 2-cut case, it is simple to numerically solve for the allowed endpoints $a_i$ using the constraints coming from $R \sim 1/\lambda$ in \ref{2cut}. This gives the green curve displayed in figure \ref{symbreak}. As we move along the curve, we are changing the filling fraction of the left and right side. The endpoints of this curve are given by the 1-cut asymmetric solution.

We could also consider 3-cut solutions. This would involve putting eigenvalues right at the maximum of the potential. Alternatively, we can think of the eigenvalue density as a charge density and the $V(x)$ as an electrostatic potential. The maximum of the potential will be a minimum if we have negative charges. So a solution with a cut near the maximum of the potential will lead to negative eigenvalue densities, which are forbidden. This agrees with the results of the bootstrap, which converge to a 1-parameter solution.

To summarize, for $g>1/4$ there is a single-cut symmetric solution. For $1/4 > g > 1/15$ there is a 2-cut solution. For $g< 1/15$ there is a one-parameter family of solutions that interpolates between an asymmetric single-cut solution and a 2-cut symmetric solution.


\section{The bootstrap approach for computing determinants or vectors}
For most of the paper, we discussed simple single-trace operators. A more complicated observable is a function of the determinant. Here we outline a strategy for computing these in the bootstrap approach.

The idea is to rewrite
\eqn{ \int dA \, {1 \over \det (z_1-A)} e^{-S(A)} = \int dA \, dv \, d \bar{v}\,e^{-S_\text{eff}(A,v) } = \mathcal{Z}_1(g,z_1) ,\\
	S_\text{eff} = S(A) + z_1 v^\dagger v - v^\dagger A v.
}We have absorbed some irrelevant factors of $(2\pi)$ into the measure. 
By adding $n$ vectors with different masses set by $z_i$, we could compute
\eqn{\ev{\prod_{i=1}^n {1 \over \det(z_i - A)}} = \mathcal{Z}_n(g, z_1, \cdots, z_n )/\mathcal{Z}_0(g). }
In general, the vectors in a matrix theory will be related to open strings: in the notation of 't Hooft, matrices are represented by double lines whereas vectors are single lines and can thus be interpreted as boundaries of the planar diagrams (for a review, see \cite{Dijkgraaf:2018vnm}). In the Liouville context, determinants in the matrix model are related to FZZT boundaries.

Now note the identity
\eqn{\pd_z \log \mathcal{Z} = - \ev{v^\dagger v}  }
This means that if we can compute correlation functions of $v$ as a function of the parameter, we can in principle reconstruct the partition function and therefore extract expectation values of determinants. Of course, other correlation functions involving vectors may be interesting in their own right, for example in large $N$ QCD, such correlation functions probe properties of the dual string.

\section{The bootstrap approach for $1/N$ corrections \la{1overN}}
In this section, we sketch how the bootstrap approach could be extended to include $1/N$ corrections. We will keep the discussion fairly theoretical; a practical discussion of how to implement the constraints and their effectiveness is left to future work.
\def\se{\succeq}
At large $N$, we defined a matrix $\M$, whose elements were single-trace expectation values. We will now adjust the notation slightly so that the entries of $\M$ are the operators $\tr \O(A,B)$ without expectation values. In other words, $\M$ will now denote a matrix of random variables instead of a matrix of expectation values. With this notation, the large $N$ constraints considered previously should be denoted $\ev{\M} \se 0$.

When we consider $1/N$ corrections, we should enforce that $\M \ge 0$, not just on expectation, but including fluctuations. The probability that $\M$ has any negative eigenvalues should always be exactly zero. One way of stating this in terms of correlation functions of $\M$ is that the generalized resolvent
\eqn{\mathcal{R}_n(x_1, \cdots x_n) = \ev{ \prod_i^n  \tr_{\!\M} {1 \over x_i-\M}   } }
is analytic in the region $x_i < 0$, for all $n$. Such expressions might look familiar from single-matrix integrals, but one should expend mental effort to keep them separate. Note that $\tr_{\!\M}$  here means a trace over correlator space, {\it not} the usual trace over $N \times N$ matrices.

A somewhat more practical statement is that if we take determinants of upper-left sub-matrices of $\M$, these are all positive random variables. Now we can ask the following question: given a sequence $m_k$, $k \in \{1,2, \cdots\}$, what are the necessary and sufficient conditions for $m_k$ to be the moments of some positive random variable?  This is known as the {\it Stieltjes moment problem}; we will now outline the solution. Define two matrices
	\def\mm{\mathfrak{M}}
	\def\hh{\mathfrak{H}}
	\def\ss{\mathfrak{S}}
\eqn{\hh &=\left[\begin{matrix}
1 & m_1 & m_2 & \cdots \\
m_1 & m_2 & m_3 & \cdots \\
m_2& m_3 & m_4 & \cdots & \\
\vdots & \vdots & \vdots & \ddots \\
\end{matrix}\right], \qquad 
	\ss =\left[\begin{matrix}
m_1 & m_2 & m_3 & \cdots & \\
m_2 & m_3 & m_4 & \cdots & \\
m_3 & m_4 & m_5 & \cdots & \\
\vdots & \vdots & \vdots & \ddots  \\
\end{matrix}\right].}
Then $m_k$ are the moments of a positive random variable if and only if $\hh \se 0$ and $\ss \se 0$. Necessity of these conditions follow from the fact that if $Y$ is any positive random variable and $P$ is an arbitrary polynomial (with complex coefficients), then
\eqn{ \ev{\bar{P}(Y) P(Y)} \ge 0, \quad \ev{Y \bar{P}(Y) P(Y) } \ge 0.}
Proving that these conditions are not just necessary but sufficient (e.g., that any sequence $m_k$ which satisfies these set of constraints corresponds to some measure on the positive reals) is more subtle; we refer the reader to \cite{simon1998classical} for an exposition. 

In our problem, we actually have a list of positive random variables $d_n$, so we are interested in the multi-variate generalization of this problem. In addition, we will in general not be able to compute all the moments of $d_n$ but only a finite list. This is known as the so-called ``truncated moment problem,'' see \cite{curto2000truncated}. 


We have so far focused on just the inequalities which follow from positivity of $\M$.
However, there are also many inequalities that just follow from the usual inequalities on moments of a collection of complex random variables. If we have a list of complex random variables $z_1 \cdots z_n$, then
\eqn{\ev{\bar{P}(\bar{z}_1, \cdots, \bar{z}_n) P(z_1 \cdots z_n)} \ge 0.}

In this case, each $z_i$ could be any single trace operator, e.g., some matrix element of $\M$. Now if we think of the monomials as basis elements of some vector space, this becomes positivity of an even bigger inner product matrix $\mm$.
\eqn{\ev{\mm} \se 0.}
Here the elements of $\mm$ are multi-trace operators, e.g., products of matrix elements of $\M$. Just as an example, in the simplest case where we only consider the constraints on a single random variable $z = \tr \O(A,B)$, the components of $\mm_{i,j} = (\tr \O)^{i+j}$.

For the single matrix model, we saw that at large $N$, the basic requirement was that the average eigenvalue density was positive. However, including $1/N$ corrections means that we are enforcing positivity of the eigenvalue distribution not just on average but including fluctuations $\rho + \delta \rho$. Even off-shell, the eigenvalue distribution must always be positive. For multi-matrix models, we know of no such simple condition.

At leading order in $1/N$, multi-trace correlators factorize, which meant that we only needed to consider loop equations of the form \nref{loopeqnmulti}. At higher orders, we need to consider more loop equations to determine the values of multi-trace correlators. In the 1-matrix model, these can be derived from
\eqn{\int dM \, \pa{}{M_{ij}} \lb \lp M^{k_1} \rp _{ij} \tr M^{k_2} \cdots \tr M^{k_n} e^{-S} \rb = 0 .}
The corresponding generalization to multi-matrices is obvious; we just consider multi-trace insertions $\tr \O_1  \tr O_2 \cdots \tr O_n$.

In the single-cut solutions to the 1-matrix model, the large $N$ eigenvalue density $\rho(\lambda)$ uniquely determines all $1/N$ corrections by topological recursion, so the above discussion is moot. However, for multi-matrix models, we do not know if this is the case. Even in the 1-matrix model, when we consider multi-cut solutions, we must impose additional requirements on the $1/N$ corrections in order to determine them uniquely, see Section 4.3 of \cite{Eynard:2015aea} for details. Said differently, does the dimension of search space increase when we include $1/N$ corrections, and if so, by how much?

\section{Ising model on a random planar lattice \la{app:ising}}
Here we outline the bootstrap approach to the Ising model on a random lattice. 
This is a 2-matrix model with interaction
\eqn{S = \tr \hf (A^2 + B^2) + {g \over 3}( A^3 + B^3) + c AB  }
For this model, we will assume $\mathbb{Z}^2$ symmetry and adopt the notation
\eqn{t_{n,m} = \tr{A^n B^m} = t_{m,n}. }
An important feature of this model is that the loop equations close on a tiny subset of all possible correlators \cite{Staudacher:1993xy}. To see this explicitly, note that for $ n \ge 1$, (a subset of) the loop equations give  \eqn{ - g t_{n+1} &= t_n + c t_{n-1, 1} - \sum_{j=0}^{n-2} t_j t_{n-2-j}  \\
 -g t_{n+1,1} &= t_{n,1} + c t_{n-1, 2} - \sum_{j=0}^{n-2} t_j t_{n-2-j,1}  \\
- g t_{n-1,2} &=  t_{n-1,1} + c t_{n} .
}
with the boundary conditions $t_0 = 1, t_{0,1} = t_1$, and  $t_{0,2} = t_2$. By substituting the third equation into the second, we can solve for all $t_n, t_{n,1}, t_{n,2}$ in terms of just $t_1$ and $t_{1,1}$.

\begin{figure}[ht]
\begin{center}
	\includegraphics[scale=.5]{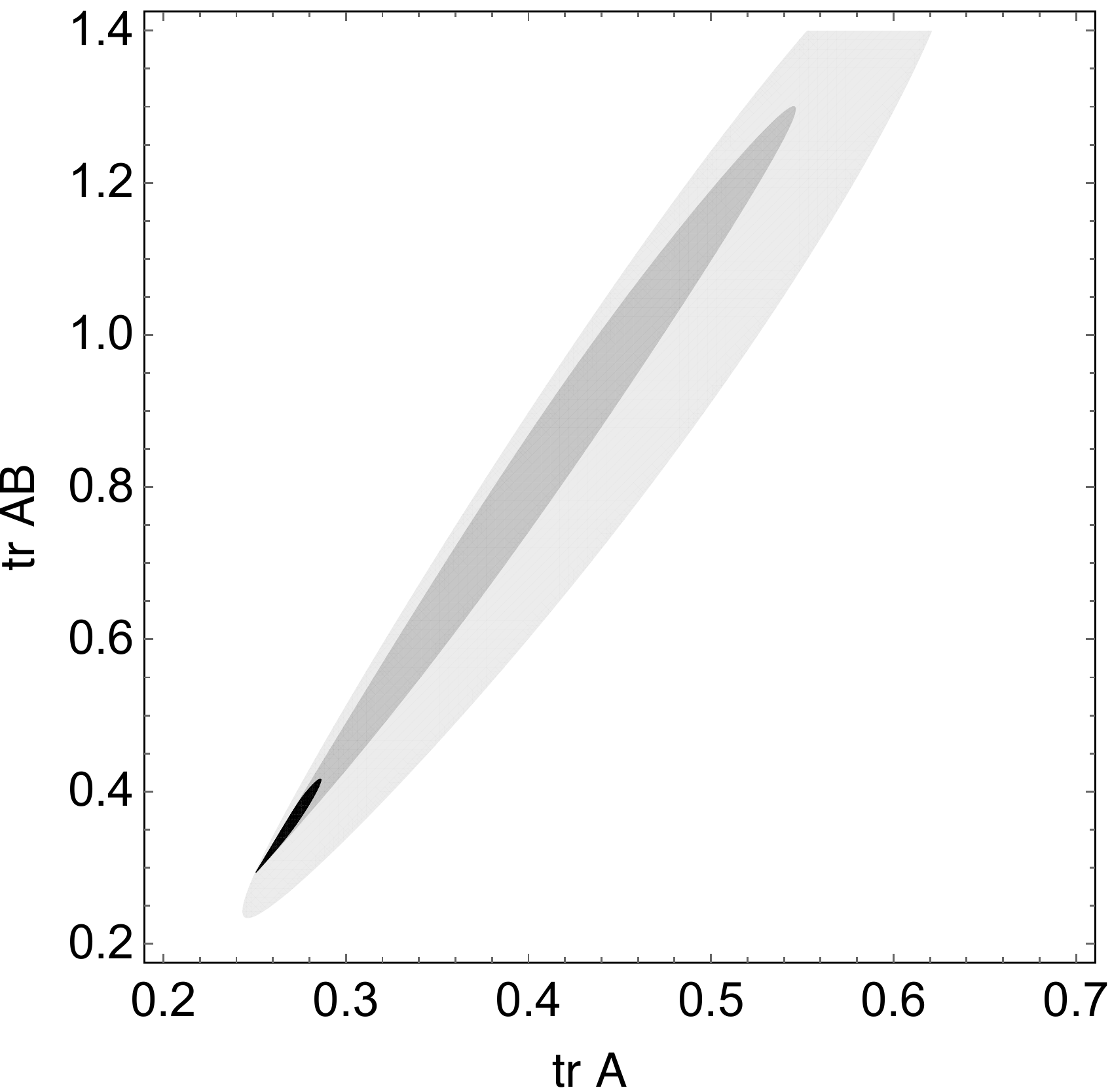}
	\caption{Constraints for the Ising model on a random planar graph with $c=g=-1/6$. Here positivity is enforced on single matrix correlators $t_k$ for inner product matrices of sizes $k = 8,12, 16$. }\label{fig:ising}
\end{center}
\end{figure}

A maximal subset of positivity constraints can be obtained by considering matrices of the form $\phi \in \text{span} \lp A^n, B A^m \rp$ for all integer $n,m \ge 0$. Then the entries of Gram matrix $\M$ will only depend on $t_k, t_{k+1}, t_{k+2}$. In practice, we found that enforcing positivity just on the single matrix correlators $t_k$ is sufficient to see convergence to the exact solution, see figure \ref{fig:ising}. This is expected from our discussion of null vectors, together with the fact that the eigenvalue distribution of a single matrix still has compact support.

For simplicity, we have focused on the Ising model with zero magnetic field where there is a $\mathbb{Z}_2$ symmetry. A simple exercise which is left to the reader is to generalize the above discussion to the Ising model in a magnetic field, in which case $g_A = g e^{h} \ne g_B = g e^{-h}$.
Another generalization is to a chain of matrices with couplings of the form $A_t A_{t+1}$. This can be thought of as a step towards solving matrix quantum mechanics, where the Euclidean time is discrete and the coupling between matrices is a discretization of the kinetic term.

\subsection{Relation to the cubic interaction \la{app:cubic}}

The matrix model defined by \ref{multi} with $W(A) = {1 \over 3} g A^3$ is related by field redefinitions to the Ising model in a magnetic field \cite{Johnston:1999he}. As we have seen the structure of the Ising model loop equations can be efficiently solved. 
Nevertheless, it is also instructive to solve the model without taking advantage of any of these simplifications which are somewhat obscured without performing the requisite field redefinitions.

\begin{figure}[ht]
\begin{center}
	\includegraphics[scale=.45]{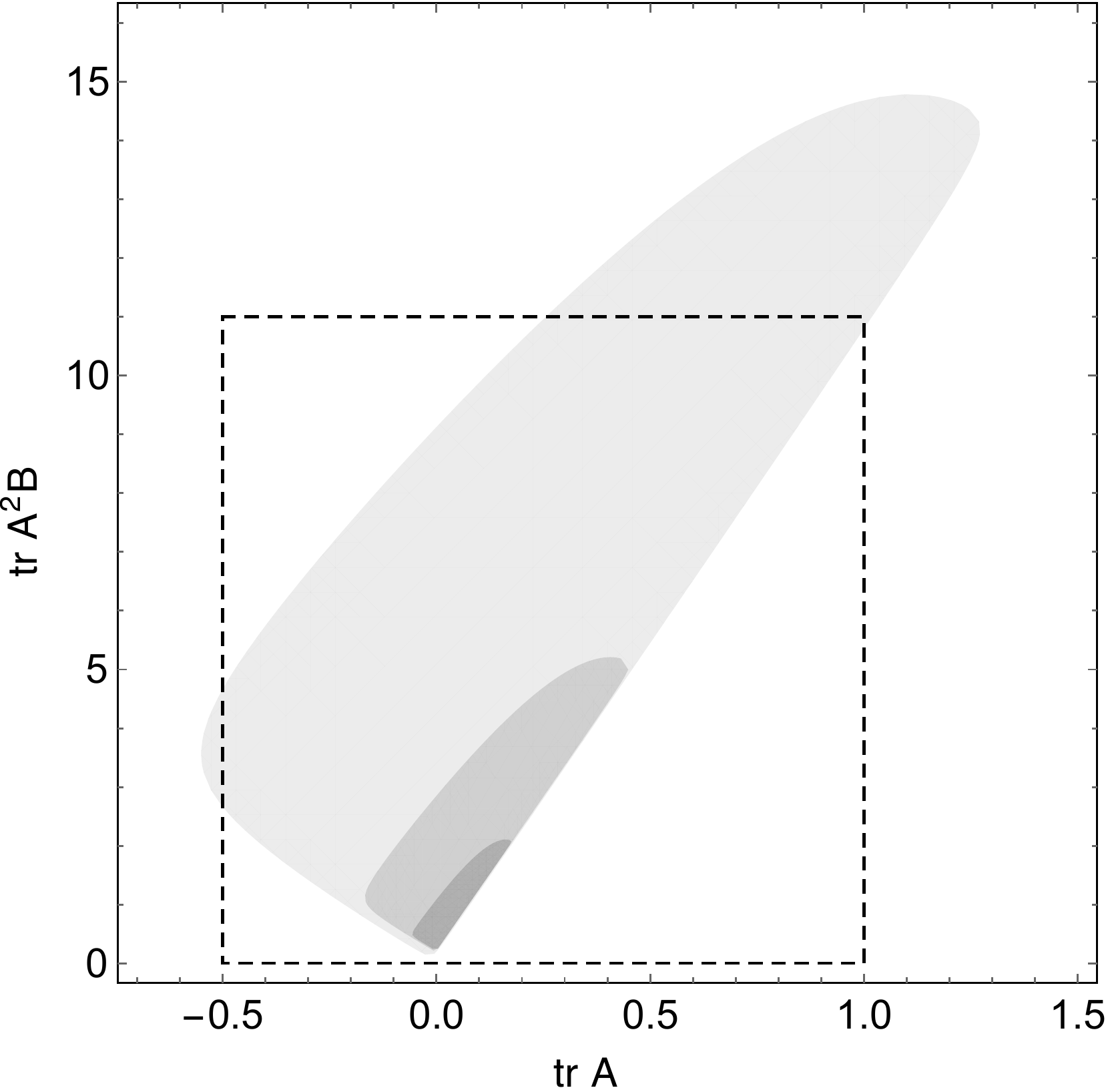} \hspace{0.2cm}
\includegraphics[scale=.45]{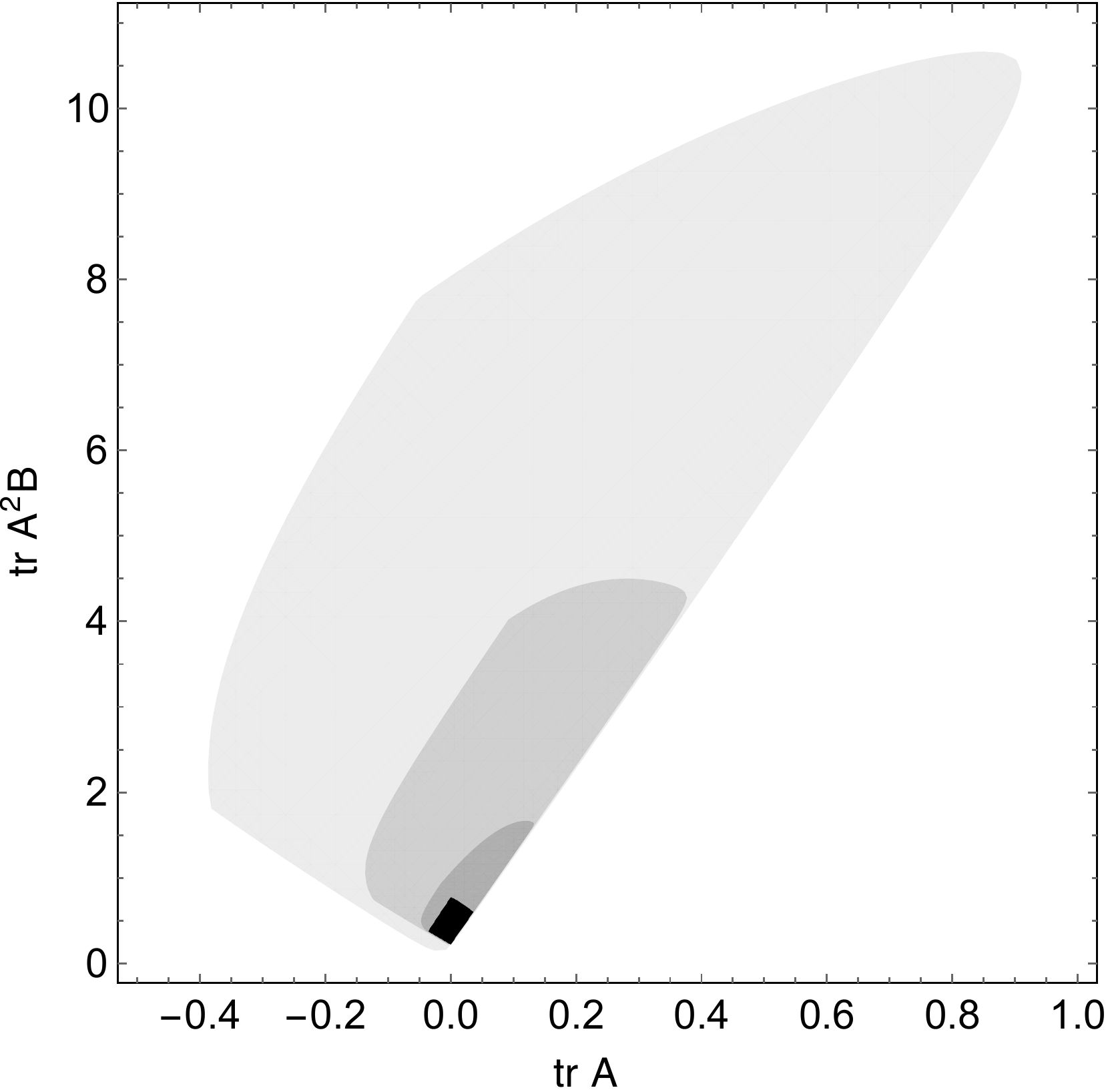}
\caption{Here we show convergence of the bootstrap as we increase the number of constraints. We chose $g = -h = 3/20 $ and show constraints from correlators of degrees up to $\{ 6,8,10\}$, corresponding to the different shades of gray. On the left, we show constraints only coming from positivity of a single matrix. On the right, we include constraints from some mixed correlators involving both $A$ and $B$ (see text for details). Imposing the full positivity constraints from all possible multi-matrix correlators of degree $\le 10$ gives the tiny black region on the right.}\label{convergence}
\end{center}
\end{figure}

We somewhat arbitrarily chose to explore the model on the line $g=-h$; we checked that studying a nearby line, e.g., $g = - 0.9 h$ does not qualitatively change our conclusions.
In figure \ref{convergence}, we show the allowed region in parameter space for $g= 3/20$. For this model, we let the search space $S$ be parameterized by $\ev{\tr A}$ and $\ev{\tr A^2 B}$. We could choose basically any two low-degree correlators and get similar results.

We study convergence as we increase the number and type of constraints. For example, we considered the constraints coming from positivity of just a single matrix. We also considered positivity constraints from ``slightly mixed'' correlators up to a fixed degree. These are correlators of the form $\ev{\tr A^{k-\ell} B^{\ell}} $. 
The advantage of such things is that the number of such correlators grows quite slowly with the degree, which means that only a rather small matrix $\M$ needs to be diagonalized. Of course, one still needs to solve the loop equations for correlators which do not enter the matrix, so we cannot get rid of the exponential complexity of the problem.
Including constraints from all possible correlators of degree $\le 10$ reduces the area of the allowed region by a factor of $\sim 4$ in this example.

\begin{figure}[ht]
\begin{center}
\includegraphics[scale=.5]{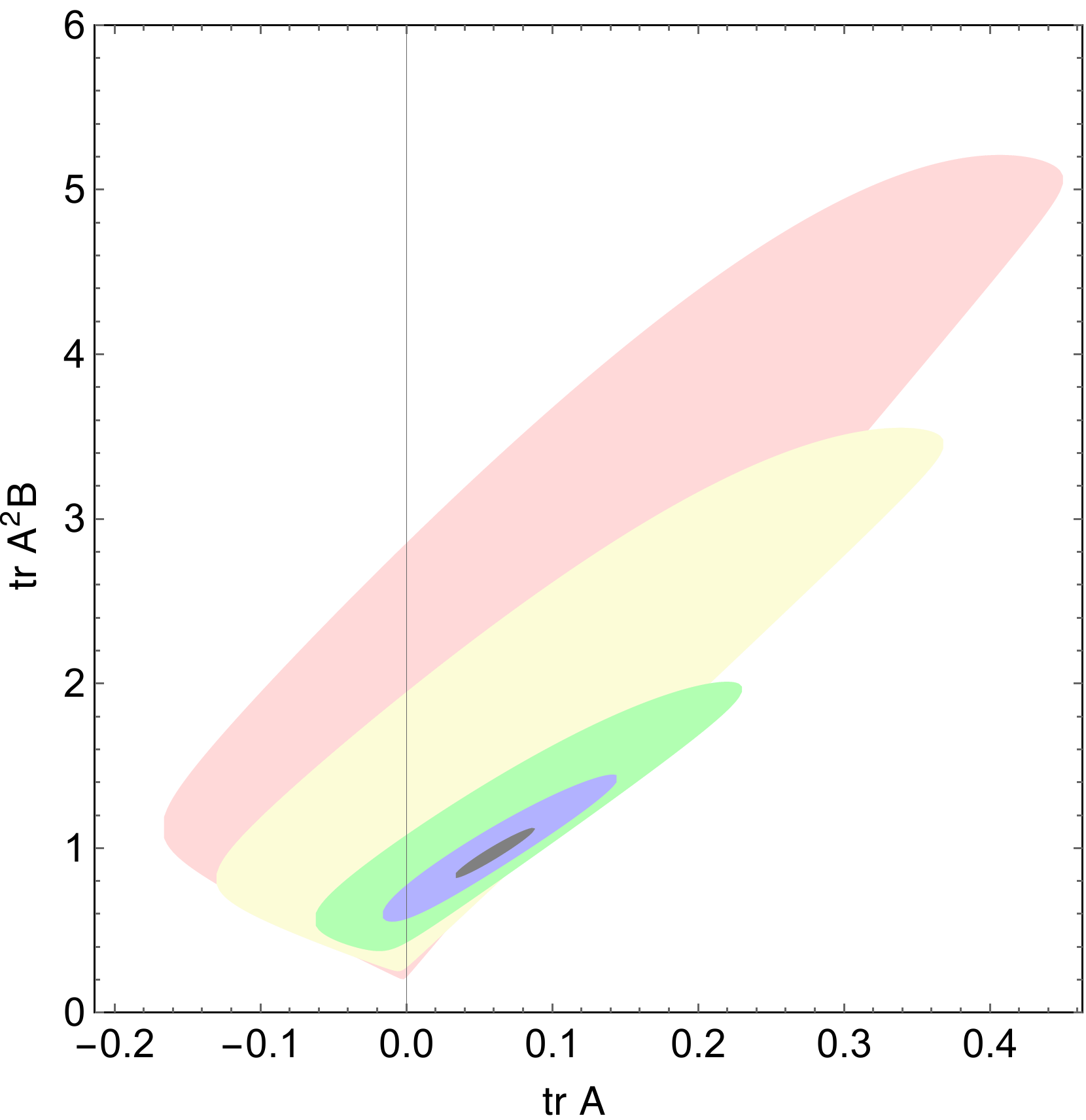}
\caption{ Here we show convergence as we approach (and pass) the critical surface in parameter space. In this plot, the number of the constraints $K$ is held fixed, but the coupling is varied. After some value of the coupling $g_e(K) = - h_e(K)$, the allowed region is empty. This gives an upper bound on the critical value of the coupling $g_* < g_e(K)$. As we increase the number of constraints $K$, we expect that $g_e(K)$ will converge to $g_*$. For visual purposes, we display constraints from single-matrix correlators of degree up to 8. This gives us the bound $g_* < .196$. Using more multi-matrix correlators up to degree 10, we find the bound $g_* < 0.185$.
}
\label{criticalpt}
\end{center}
\end{figure}

We searched for a critical point on the $g=-h$ line. By using all the constraints up to degree 10, we were able to bound the critical point $g_* = -h_* < 0.185$. This implies that there is also a critical point at $g = -h_* > -0.185$.
In figure \ref{criticalpt}, we show convergence as we approach this critical point. In this figure, we do {\it not} vary the constraints that are being checked; instead we are changing the coupling. 
This figure should be viewed as the higher dimensional analog of the ``peninsula'' depicted in figure \ref{peninsula}.
In general, we can use this method to constrain the allowed region in the 4-dimensional space parameterized by $\ev{\tr A}, \ev{\tr A^2 B}, g,h$.

\section{Mathematica code for generating loop equations}
For the convenience of the reader, a Mathematica code is provided in the source directory of this paper that may be used to generate the large $N$ loop equations for single trace correlation functions.
The code may be easily modified to derive the loop equations of any multi-matrix model with arbitrary polynomial interactions. 
In Mathematica notation, a string $\{1,2,1,1,2\}$ corresponds to a string of matrices $ABAAB$. We use the notation $e_{\{1,2,1,1,2\}} = \tr ABAAB$.


\mciteSetMidEndSepPunct{}{\ifmciteBstWouldAddEndPunct.\else\fi}{\relax}
\bibliographystyle{utphys}
\bibliography{BigReferencesFile.bib}{}

\providecommand{\href}[2]{#2}\begingroup\raggedright\begin{thebibliography}{10}

\bibitem{tHooft:1973jz}
G.~'t~Hooft, ``{A Planar Diagram Theory for Strong Interactions},''
\href{http://dx.doi.org/10.1016/0550-3213(74)90154-0}{{\em Nucl. Phys.}
  {\bfseries B72} (1974) 461}.

\bibitem{'tHooft:1974hx}
G.~'t~Hooft, ``{A Two-Dimensional Model for Mesons},''
\href{http://dx.doi.org/10.1016/0550-3213(74)90088-1}{{\em Nucl. Phys.}
  {\bfseries B75} (1974) 461--470}.

\bibitem{wigner1993characteristic}
E.~P. Wigner, ``Characteristic vectors of bordered matrices with infinite
  dimensions i,'' in {\em The Collected Works of Eugene Paul Wigner},
  pp.~524--540.
\newblock Springer, 1955.

\bibitem{bipz}
E.~Br{\'e}zin, C.~Itzykson, G.~Parisi, and J.-B. Zuber, ``Planar diagrams,'' in
  {\em The Large N Expansion In Quantum Field Theory And Statistical Physics:
  From Spin Systems to 2-Dimensional Gravity}, pp.~567--583.
\newblock World Scientific, 1993.

\bibitem{Bessis:1980ss}
D.~Bessis, C.~Itzykson, and J.~B. Zuber, ``{Quantum field theory techniques in
  graphical enumeration},''
\href{http://dx.doi.org/10.1016/0196-8858(80)90008-1}{{\em Adv. Appl. Math.}
  {\bfseries 1} (1980) 109--157}.

\bibitem{douglas1990strings}
M.~R. Douglas and S.~H. Shenker, ``Strings in less than one dimension,'' {\em
  Nuclear Physics B} {\bfseries 335} no.~3, (1990) 635--654.

\bibitem{Kazakov:2000aq}
V.~A. Kazakov, ``{Solvable matrix models},''
\newblock 2000.
\newblock
\href{http://arxiv.org/abs/hep-th/0003064}{{\ttfamily arXiv:hep-th/0003064
  [hep-th]}}.
\newblock

\bibitem{eynard2011formal}
B.~Eynard, ``Formal matrix integrals and combinatorics of maps,'' in {\em
  Random matrices, random processes and integrable systems}, pp.~415--442.
\newblock Springer, 2011.

\bibitem{Klebanov:2003wg}
I.~R. Klebanov, J.~M. Maldacena, and N.~Seiberg, ``{Unitary and complex matrix
  models as 1-d type 0 strings},''
  \href{http://dx.doi.org/10.1007/s00220-004-1183-7}{{\em Commun. Math. Phys.}
  {\bfseries 252} (2004) 275--323},
\href{http://arxiv.org/abs/hep-th/0309168}{{\ttfamily arXiv:hep-th/0309168
  [hep-th]}}.

\bibitem{DiFrancesco:1993cyw}
P.~Di~Francesco, P.~H. Ginsparg, and J.~Zinn-Justin, ``{2-D Gravity and random
  matrices},'' \href{http://dx.doi.org/10.1016/0370-1573(94)00084-G}{{\em Phys.
  Rept.} {\bfseries 254} (1995) 1--133},
\href{http://arxiv.org/abs/hep-th/9306153}{{\ttfamily arXiv:hep-th/9306153
  [hep-th]}}.

\bibitem{Anderson:2016rcw}
P.~D. Anderson and M.~Kruczenski, ``{Loop Equations and bootstrap methods in
  the lattice},'' \href{http://dx.doi.org/10.1016/j.nuclphysb.2017.06.009}{{\em
  Nucl. Phys.} {\bfseries B921} (2017) 702--726},
\href{http://arxiv.org/abs/1612.08140}{{\ttfamily arXiv:1612.08140 [hep-th]}}.

\bibitem{Jevicki:1983hb}
A.~Jevicki and J.~P. Rodrigues, ``{Master Variables and Spectrum Equations in
  Large $N$ Theories},''
  \href{http://dx.doi.org/10.1016/0550-3213(84)90216-5}{{\em Nucl. Phys. B}
  {\bfseries 230} (1984) 317--335}.

\bibitem{curto2000truncated}
R.~Curto and L.~Fialkow, ``The truncated complex k-moment problem,'' {\em
  Transactions of the American mathematical society} {\bfseries 352} no.~6,
  (2000) 2825--2855.

\bibitem{burgdorf2012truncated}
S.~Burgdorf and I.~Klep, ``The truncated tracial moment problem,'' {\em Journal
  of Operator Theory} (2012) 141--163.

\bibitem{Itzykson:1979fi}
C.~Itzykson and J.~B. Zuber, ``{The Planar Approximation. 2.},''
\href{http://dx.doi.org/10.1063/1.524438}{{\em J. Math. Phys.} {\bfseries 21}
  (1980) 411}.

\bibitem{ikkt}
N.~Ishibashi, H.~Kawai, Y.~Kitazawa, and A.~Tsuchiya, ``{A Large N reduced
  model as superstring},''
  \href{http://dx.doi.org/10.1016/S0550-3213(97)00290-3}{{\em Nucl. Phys.}
  {\bfseries B498} (1997) 467--491},
\href{http://arxiv.org/abs/hep-th/9612115}{{\ttfamily arXiv:hep-th/9612115
  [hep-th]}}.

\bibitem{Klebanov:1991qa}
I.~R. Klebanov, ``{String theory in two-dimensions},'' in {\em {Spring School
  on String Theory and Quantum Gravity (to be followed by Workshop) Trieste,
  Italy, April 15-23, 1991}}, pp.~30--101.
\newblock 1991.
\newblock
\href{http://arxiv.org/abs/hep-th/9108019}{{\ttfamily arXiv:hep-th/9108019
  [hep-th]}}.
\newblock

\bibitem{Klebanov:2018fzb}
I.~R. Klebanov, F.~Popov, and G.~Tarnopolsky, ``{TASI Lectures on Large $N$
  Tensor Models},'' \href{http://dx.doi.org/10.22323/1.305.0004}{{\em PoS}
  {\bfseries TASI2017} (2018) 004},
\href{http://arxiv.org/abs/1808.09434}{{\ttfamily arXiv:1808.09434 [hep-th]}}.

\bibitem{Eynard:2015aea}
B.~Eynard, T.~Kimura, and S.~Ribault, ``{Random matrices},''
\href{http://arxiv.org/abs/1510.04430}{{\ttfamily arXiv:1510.04430 [math-ph]}}.

\bibitem{Cicuta:1986pu}
G.~M. Cicuta, L.~Molinari, and E.~Montaldi, ``{Large $N$ Phase Transitions in
  Low Dimensions},''
\href{http://dx.doi.org/10.1142/S021773238600018X}{{\em Mod. Phys. Lett.}
  {\bfseries A1} (1986) 125}.

\bibitem{Dijkgraaf:2018vnm}
R.~Dijkgraaf and E.~Witten, ``{Developments in Topological Gravity},''
  \href{http://dx.doi.org/10.1142/S0217751X18300296}{{\em Int. J. Mod. Phys.}
  {\bfseries A33} no.~30, (2018) 1830029},
\href{http://arxiv.org/abs/1804.03275}{{\ttfamily arXiv:1804.03275 [hep-th]}}.

\bibitem{simon1998classical}
B.~Simon, ``The classical moment problem as a self-adjoint finite difference
  operator,'' {\em Advances in Mathematics} {\bfseries 137} no.~1, (1998)
  82--203.

\bibitem{Staudacher:1993xy}
M.~Staudacher, ``{Combinatorial solution of the two matrix model},''
  \href{http://dx.doi.org/10.1016/0370-2693(93)91063-S}{{\em Phys. Lett.}
  {\bfseries B305} (1993) 332--338},
\href{http://arxiv.org/abs/hep-th/9301038}{{\ttfamily arXiv:hep-th/9301038
  [hep-th]}}.

\bibitem{Johnston:1999he}
D.~Johnston, ``{Symmetric vertex models on planar random graphs},''
  \href{http://dx.doi.org/10.1016/S0370-2693(99)00948-X}{{\em Phys. Lett. B}
  {\bfseries 463} (1999) 9--18},
  \href{http://arxiv.org/abs/cond-mat/9812169}{{\ttfamily
  arXiv:cond-mat/9812169}}.

\end{thebibliography}\endgroup

\end{document}